\documentclass{article}

\usepackage{arxiv}

\usepackage[utf8]{inputenc}
\usepackage[T1]{fontenc}
\usepackage{hyperref}
\usepackage{url}
\usepackage{booktabs}
\usepackage{amsfonts}
\usepackage{nicefrac}
\usepackage{microtype}
\usepackage{graphicx}
\usepackage{cite}
\usepackage{tabularx}
\usepackage{listings}
\usepackage{placeins}
\usepackage{caption}
\usepackage{float}
\usepackage{doi}
\usepackage{colortbl}
\usepackage[table,xcdraw]{xcolor}

\title{Scalable APT Malware Classification via Parallel Feature Extraction and GPU-Accelerated Learning}

\renewcommand{\shorttitle}{Scalable APT Malware Classification via Parallel Feature Extraction and GPU-Accelerated Learning}

\hypersetup{
  pdftitle={Scalable APT Malware Classification},
  pdfauthor={Noah Subedar, Taeui Kim, Saathwick Venkataramalingam},
  pdfkeywords={Malware Classification, Automation, Parallel Processing, Machine Learning, Neural Network Models, GPU-Accelerated Computing},
}

\begin{document}
\definecolor{bggray}{rgb}{0.85,0.85,0.85} %
\definecolor{OliveGreen}{rgb}{0.33, 0.42, 0.18}

\lstdefinestyle{bashStyle}{
  backgroundcolor=\color{gray!10},
  basicstyle=\ttfamily\small,
  frame=none,
  keywordstyle=\color{blue},
  commentstyle=\color{OliveGreen},
  stringstyle=\color{red},
  breaklines=true,
  showstringspaces=false,
  showlines=true,
}

\author{
  \href{https://orcid.org/0009-0004-5437-2325}{\includegraphics[scale=0.06]{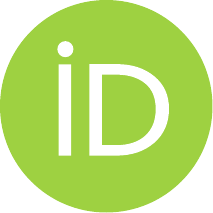}\hspace{1mm}Noah Subedar} \\
  Master of Cybersecurity and Threat Intelligence \\
  University of Guelph \\
  HBSc, Computer Science Specialist \\
  University of Toronto \\
  \texttt{nsubedar@uoguelph.ca} \\\\
  \begin{tabular}{cc}
    \begin{tabular}{c}
      \href{https://orcid.org/0009-0003-9772-6886}{\includegraphics[scale=0.06]{orcid.pdf}\hspace{1mm}Taeui Kim} \\
      Master of Cybersecurity and Threat Intelligence \\
      University of Guelph \\
      BSc, Software Engineering \\
      University of New Brunswick \\
      \texttt{taeui@uoguelph.ca}
    \end{tabular}
    &
    \begin{tabular}{c}
      \href{https://orcid.org/0009-0007-2025-826X}{\includegraphics[scale=0.06]{orcid.pdf}\hspace{1mm}Saathwick Venkataramalingam} \\
      Master of Cybersecurity and Threat Intelligence \\
      University of Guelph \\
      Post-Graduate Certificate in Offensive Cyber Security \\
      York University \\
      \texttt{venkatas@uoguelph.ca}
    \end{tabular}
  \end{tabular}
}

\maketitle

\begin{abstract}

This paper presents an underlying framework for both automating and accelerating malware classification, more specifically, mapping malicious executables to known Advanced Persistent Threat (APT) groups. The main feature of this analysis is the assembly-level instructions present in executables which are also known as opcodes. The collection of such opcodes on many malicious samples is a lengthy process; hence, open-source reverse engineering tools are used in tandem with scripts that leverage parallel computing to analyze multiple files at once. Traditional and deep learning models are applied to create models capable of classifying malware samples. One-gram and two-gram datasets are constructed and used to train models such as SVM, KNN, and Decision Tree; however, they struggle to provide adequate results without relying on metadata to support n-gram sequences. The computational limitations of such models are overcome with convolutional neural networks (CNNs) and heavily accelerated using graphical compute unit (GPU) resources.

\end{abstract}


\section{Introduction}
The rapid evolution of malware, particularly those associated with APT groups, present a significant challenge to cybersecurity researchers and analysts \cite{1, 2}. Traditional signature-based detection techniques often fail to recognize novel or obfuscated threats, necessitating more robust, data-driven approaches to malware classification. One promising avenue involves the analysis of opcode sequences which represent low-level machine instructions extracted from malicious binaries. These sequences capture behavioural patterns that are difficult to mask, even under obfuscation or packing; however, extracting and analyzing large-scale opcode datasets remains computationally intensive and prone to bottlenecks when conducted sequentially or with manual intervention\cite{1h,3}.

To address these limitations, this study introduces a fully automated, parallelized pipeline for malware classification based on opcode analysis, encompassing traditional machine learning and deep learning techniques. By leveraging open-source tools like Ghidra in headless mode, combined with Python scripts, thousands of malicious executables are efficiently decompiled and transformed into structured opcode datasets \cite{2h,5}. These are further processed into n-gram representations and fed into models such as Support Vector Machines (SVM), K-Nearest Neighbors (KNN), Decision Trees, and GPU-accelerated Convolutional Neural Networks (CNNs)\cite{67}. This comprehensive approach not only streamlines the data extraction and training process but also highlights the performance trade-offs across various classifiers, metadata usage, and dataset configurations including one-to-one and one-to-many label mappings\cite{3h,4,6}.

\section{Scripts \& Environment Setup}
\subsection{Container Environment}
To facilitate ease of use and reproducibility, a Podman container comprising the scripts and tools discussed in this paper has been published on Docker Hub. It is recommended to use Podman along with a NVIDIA GPU to run the container with hardware acceleration. The following instructions demonstrate how to run the container on a Fedora Linux system which includes Podman by default. They also outline the steps required to enable support for NVIDIA GPUs, assuming the necessary drivers have been installed correctly.

\subsubsection{Enabling NVIDIA GPU Support}
To enable the container to interface with NVIDIA GPUs, the NVIDIA Container Toolkit must be installed, and a Container Device Interface (CDI) specification file must be generated to expose the GPUs to the container environment \cite{podmanGPU}.
\begin{lstlisting}[language=bash, style=bashStyle]
sudo dnf install nvidia-container-toolkit
sudo nvidia-ctk cdi generate --output=/etc/cdi/nvidia.yaml
\end{lstlisting}
Fedora Linux enables Security-Enhanced Linux (SELinux) by default which restricts containers from accessing host devices for security reasons. To permit containers to use such devices, SELinux must be configured accordingly using the following command \cite{podmanGPU}:
\begin{lstlisting}[language=bash, style=bashStyle]
sudo setsebool -P container_use_devices true
\end{lstlisting}

\subsubsection{Data Format}
Before running the container, it is crucial that the data provided to it is formatted correctly. The scripts discussed in this paper require a specific file structure and naming scheme. The organizational structure is outlined below: the left side shows the required directory structure, and the right side presents an example of the data formatted correctly.

\begin{minipage}{0.5\textwidth}
\begin{lstlisting}[language=bash, style=bashStyle]
data
|-- APT Group
    |-- Software Name
        |-- Malicious Executable








        
\end{lstlisting}
\end{minipage}%
\begin{minipage}{0.5\textwidth}
\begin{lstlisting}[language=bash, style=bashStyle]
data
|-- G0001
|   |-- Software A
|   |   |-- G0001_Malware1.exe
|   |   `-- G0001_Malware2.dll
|   `-- Software B
|       `-- G0001_Malware3.exe
|-- G0002
|   |-- G0002_Malware4.exe
|   `-- Software C
|       `-- G0002_Malware5.exe
`-- G0003
    `-- G0003_Malware6.exe
\end{lstlisting}
\end{minipage}

\subsubsection{Running the Container}
To run the container, first pull it from Docker Hub and execute it in interactive mode with a bash shell. Within the container, the shell script named \texttt{run.sh} can be run to automatically execute all methods outlined in the paper with results saved to \texttt{/app/results} directory within the container. By default, all scripts are executed with the arguments specified in the paper; however, these can be customized by modifying the environment variables, editing \texttt{run.sh} or manually running the individual scripts. To provide malware samples to the container, mount a host directory containing the structure defined above to the container's \texttt{/data} directory.

\begin{lstlisting}[language=bash, style=bashStyle]
# Host
podman run --rm -it \
  --device nvidia.com/gpu=all \
  -v /path/to/host/directory:/data:z \
  docker.io/noahsub/scalable-apt-malware-classification

# Container
./run.sh
\end{lstlisting}

\subsection{Scripts}
Although the container offers an automated method for analyzing malware samples, the scripts are also available on the paper's associated GitHub page \cite{github}, specifically in the releases section, along with the collected malware samples. The functionality and usage of each script are defined below.

\subsubsection{Ghidra Manager}
The \texttt{ghidra\_manager.py} script is responsible for managing the extraction of opcodes from malicious executables using Ghidra headless mode.
\begin{lstlisting}[language=bash, style=bashStyle]
python3 ghidra_manager.py \
    --GHIDRA <path_to_Ghidra_headless_analyzer> \
    --DIRECTORY <path_to_malware_root_folder> \
    --THREADS <number_of_threads> \
    --SKIP <true|false> \
    --TIMEOUT 1200
\end{lstlisting}

\begin{table}[H]
  \renewcommand{\arraystretch}{1.3}
  \centering
  \caption{Ghidra Manager Command Line Flags Description}
  \begin{tabularx}{\textwidth}{|l|X|c|c|}
    \hline
    \rowcolor[rgb]{0.85, 0.85, 0.85} \textbf{Flag} & \textbf{Description} & \textbf{Required} & \textbf{Default Value} \\
    \hline
    \rowcolor[rgb]{0.95, 0.95, 0.95} \texttt{--GHIDRA} & The path to Ghidra’s headless analyzer. & Yes & N/A \\
    \hline
    \texttt{--DIRECTORY} & The path to the directory containing malware. & Yes & N/A \\
    \hline
    \rowcolor[rgb]{0.95, 0.95, 0.95} \texttt{--THREADS} & The number of threads to use, i.e., how much malware to analyze at once. & No & 4 \\
    \hline
    \texttt{--SKIP} & If a malicious executable already has had its opcodes extracted then skip analyzing it. & No & No \\
    \hline
    \rowcolor[rgb]{0.95, 0.95, 0.95} \texttt{--TIMEOUT} & The maximum amount of time given to analyze a single malicious executable. & No & 1200 \\
    \hline
  \end{tabularx}
  \label{tab:command_flags}
\end{table}

\subsubsection{Preprocess}
The \texttt{preprocess.py} script creates an optimized dataset suitable for machine learning models using opcodes extracted from collected malware samples.
\begin{lstlisting}[language=bash, style=bashStyle]
python3 preprocess.py \
    --opcodes <path_to_directory_containing_opcodes> \
    -n 2 \
    --percentiles <percentile_1>,<percentile_n>
\end{lstlisting}
\begin{table}[H]
  \renewcommand{\arraystretch}{1.3} %
  \centering
  \caption{Preprocess Command Line Flags Description}
  \begin{tabularx}{\textwidth}{|>{\raggedright\arraybackslash}X|>{\raggedright\arraybackslash}X|>{\raggedright\arraybackslash}X|>{\raggedright\arraybackslash}X|}
    \hline
    \rowcolor[rgb]{0.85, 0.85, 0.85} \textbf{Argument} & \textbf{Usage} & \textbf{Required} & \textbf{Default Value} \\
    \hline
    \rowcolor[rgb]{0.95, 0.95, 0.95} \texttt{--opcodes} & Specifies the path to the directory containing the .opcode files. & Yes & N/A \\
    \hline
    \texttt{-n} & Sets the maximum n-gram size & Yes & 2 \\
    \hline
    \rowcolor[rgb]{0.95, 0.95, 0.95} \texttt{--percentiles} & Accepts a comma-separated list of percentiles. & Yes & N/A \\
    \hline
  \end{tabularx}
  \label{tab:argument_descriptions}
\end{table}

\subsubsection{Classifier}
The \texttt{classifier.py} script classifies malicious samples by using opcodes and metadata with three types of classifiers: SVM, KNN, and Decision Tree.

\begin{lstlisting}[language=bash, style=bashStyle]
python3 classifier.py --dataset <path_to_csv_or_pkl_dataset>
\end{lstlisting}

\begin{table}[H]
  \renewcommand{\arraystretch}{1.3} %
  \centering
  \caption{Classifier Command Line Flags Description}
  \begin{tabularx}{\textwidth}{|>{\raggedright\arraybackslash}X|>{\raggedright\arraybackslash}X|>{\raggedright\arraybackslash}X|}
    \hline
    \rowcolor[rgb]{0.85, 0.85, 0.85} \textbf{Argument} & \textbf{Usage} & \textbf{Required} \\
    \hline
    \rowcolor[rgb]{0.95, 0.95, 0.95} \texttt{--dataset} & The dataset to use for training and performance evaluation. Can be in \texttt{.csv} or \texttt{.plk} (serialized pandas dataframe) format. & Yes \\
    \hline
  \end{tabularx}
  \label{tab:model_arguments}
\end{table}

\subsubsection{CNN Preprocess}
The \texttt{preprocess-cnn.py} script intelligently removes samples from the dataset that cause one-to-many mappings.

\begin{lstlisting}[language=bash, style=bashStyle]
python3 preprocess.py --dataset <dataset_name>
\end{lstlisting}

\begin{table}[H]
  \centering
  \caption{CNN Preprocessing Command Line Flags Description}
  \label{tab:preprocessing-args}
  \renewcommand{\arraystretch}{1.3}
  \begin{tabularx}{\textwidth}{|l|X|}
    \hline
    \rowcolor[gray]{0.85} \textbf{Argument} & \textbf{Usage} \\
    \hline
    \rowcolor[gray]{0.95} \texttt{--dataset} & The path to the dataset of opcodes. \\
    \hline
  \end{tabularx}
\end{table}

\subsubsection{CNN Model}
The \texttt{model.py} script trains a CNN model to classify malware based on opcode sequences inspired by the paper \textit{Deep Android Malware Detection} \cite{deep}.

\begin{lstlisting}[language=bash, style=bashStyle]
python3 model.py \
    --directory <data_directory> \
    --percentile <percentile_value> \
    --k <k_value> \
    --epochs <num_epochs> \
    --batch_size <batch_size> \
    --validation_split <validation_fraction>
\end{lstlisting}

\begin{table}[H]
  \renewcommand{\arraystretch}{1.3} %
  \centering
  \caption{Model Script Arguments}
  \begin{tabularx}{\textwidth}{|>{\raggedright\arraybackslash}l|>{\raggedright\arraybackslash}X|>{\raggedright\arraybackslash}X|>{\raggedright\arraybackslash}X|}
    \hline
    \rowcolor[rgb]{0.85, 0.85, 0.85} \textbf{Argument} & \textbf{Usage} & \textbf{Required} & \textbf{Default Value} \\
    \hline
    \rowcolor[rgb]{0.95, 0.95, 0.95} \texttt{--directory} & The directory containing the opcode files. & Yes & N/A \\
    \hline
    \texttt{--percentile} & The percentile to determine the maximum sequence length. & No & 50 \\
    \hline
    \rowcolor[rgb]{0.95, 0.95, 0.95} \texttt{--k} & The output dimension of the embedding layer. & No & 8 \\
    \hline
    \texttt{--epochs} & The number of epochs to train the model. & No & 16 \\
    \hline
    \rowcolor[rgb]{0.95, 0.95, 0.95} \texttt{--batch\_size} & The batch size to train the model. & No & 32 \\
    \hline
    \texttt{--validation\_split} & The validation split to use during training. & No & 0.1 \\
    \hline
  \end{tabularx}
  \label{tab:model_arguments}
\end{table}

\section{Methodology: Opcode Extraction}
The extraction of machine-level instructions (opcodes) from malicious executables were automated using Ghidra’s headless mode in conjunction with Python scripting. This approach optimizes the creation and management of Ghidra projects while enabling concurrent opcode extraction through multiple processes, significantly enhancing efficiency in malware analysis and reverse engineering.

\subsection{Sample Collection}
Malicious executable samples linked to various APT groups and their associated software were gathered from multiple malware databases. The collected samples have been made available through the paper's GitHub repository \cite{github}.

\subsection{Engine Selection}
Before extracting the opcodes from collected malicious executable samples, the engine to be used as the base for the scripts had to be determined. Various options were explored including IDA Pro, Ghidra, and Binary Ninja. Each option had its advantages and disadvantages. Both IDA Pro and Binary Ninja were relatively easy to work with with Binary Ninja being heavily focused on scripting which made it ideal for the analysis of executable malware \cite{7,8}. Ultimately, Ghidra was selected due to its open-source nature and ability to operate in a headless environment \cite{rohleder2019hands}.

\subsection{Automated Scripts}
Ghidra's headless analyzer allowed an executable and a \texttt{Python2} script to be passed in as arguments. This enabled Ghidra to automatically analyze the provided executable and then run the provided script \cite{markarian2023function}. In this case, the script's purpose was to extract opcodes.

The automation framework was structured into two primary components:
\begin{itemize} 
    \item \textbf{Opcode Extraction Component} – Analyzed each binary, retrieved machine-level instructions, and extracted the corresponding opcode sequences. 
    \item \textbf{Manager Component} – Managed process spawning, file processing, and orchestrated Ghidra’s functionality, ensuring seamless execution of the analysis workflow. 
\end{itemize}

Given the large volume of malicious executables requiring analysis, efficiency was identified as a critical priority. To address this, parallel computing was leveraged, enabling simultaneous extraction of opcodes from multiple samples. This approach was essential, as the dataset comprised nearly ten gigabytes of executable malware—totaling 4,630 individual samples \cite{9, 10}.

The automated opcode extraction process followed these structured steps:

\begin{enumerate}
    \item \textbf{Identification of Executables} – Located all executable files within the specified directory containing malware samples. This included malware in subdirectories of the specified directory.
    \item \textbf{Project Initialization} – Created a separate Ghidra project for each binary to ensure an isolated and organized analysis environment.
    \item \textbf{Binary Importation} – Loaded the malware executable into its corresponding Ghidra project.
    \item \textbf{Opcode Extraction} – Executed \texttt{oopcode\_extractor.py} within Ghidra’s headless mode to extract machine-level opcodes.
    \item \textbf{Storage and Organization} – Saved extracted opcode sequences in a designated output directory for further analysis.
    \item \textbf{Performance Optimizations} – Implemented safeguards such as timeout handling and file tracking to enhance efficiency and prevent redundant processing.
    \item \textbf{Cleanup Operations} – Removed unnecessary files to maintain a streamlined and organized workspace.
\end{enumerate}

By automating opcode extraction using Ghidra and Python, this methodology improved scalability, reduced manual intervention, and enhanced the overall efficiency of malware analysis workflows.

\subsection{Hardware Utilized}
To run the scripts, an isolated Linux machine was utilized with the following configuration:

\begin{table}[H]
\centering
\renewcommand{\arraystretch}{1.4}
\caption{System Hardware Specifications Used for Opcode Extraction}
\label{tab:hardware_specs}
\begin{tabularx}{\textwidth}{|l|l|X|}
\hline
\rowcolor[rgb]{0.85, 0.85, 0.85}
\textbf{Hardware Type} & \textbf{Model} & \textbf{Specifications} \\
\hline
\rowcolor[rgb]{0.95, 0.95, 0.95}
CPU & AMD Ryzen 7 3700X & 8 cores / 16 threads, Base Clock: 3.6GHz, Boost Clock: up to 4.4GHz \\
\hline
GPU & NVIDIA GeForce RTX 3060 & 12GB GDDR6 VRAM, 3584 CUDA cores, Boost Clock: 1.78GHz \\
\hline
\rowcolor[rgb]{0.95, 0.95, 0.95}
Memory & DDR4 Non-ECC & 16GB, 3200MHz \\
\hline
Operating System & Pop!\_OS & Version 22.04 LTS (64-bit) \\
\hline
\end{tabularx}
\end{table}

\section{Methodology: N-Gram Dataset \& Training}

\subsection{Preprocessing}
To utilize the raw extracted opcodes, they first had to be preprocessed into a usable format. The preprocessing pipeline begins by recursively discovering all .opcode files in the provided directory. This process stores the relative path from the specified directory for each of the files which contains metadata such as the associated group and malware name; the metadata was based on the diligent organization structure utilized during opcode extraction \cite{11,12}. To ensure consistency in machine learning models, all opcodes were converted to uppercase, preventing variations in case from being misinterpreted as distinct instructions. Before generating n-gram sequences, the extracted opcode lines had to be adjusted to ensure their length was divisible by the selected n-gram size. This was achieved by padding the extracted opcode lines with “PAD” tokens until they reached the required length.

\subsection{N-Gram Sequence Generation}

The opcodes were then grouped into n-gram sequences:
\begin{itemize}
    \item \textbf{1-Gram (Unigram)}: Each opcode was treated individually capturing its frequency across the sample.
    \item \textbf{2-Gram (Bigram)}: Pairs of consecutive opcodes were generated capturing local context and instruction transitions.
\end{itemize}

\subsection{Vocabulary Generation}
Although each opcode file contained a different set of instructions (unless they were the same malware), the datasets used for training and testing must contain each unique opcode or n-gram opcode pair so that they can store their frequencies across all files. To achieve this, all opcode files were iterated over the generated n-gram sequences which were then converted to sets to remove duplicates. This effectively allowed the creation of headers (features) for the dataset.

\subsection{Dataset Construction}

\subsubsection{Extracting Metadata}
In addition to opcode features stored in the dataset, relevant metadata was extracted to serve as labels. This metadata included the associated APT group, the name of the malware, and the type of the executable. The metadata was extracted from the relative path of the opcode file which contained such information. This step was carefully planned during the opcode extraction phase to ensure that metadata was retained even after converting the malware into \texttt{.opcode} files. If any given piece of metadata was missing for an opcode file, it was simply listed as unknown in the dataset.

\subsubsection{Generating Feature Data}
For each malware sample, a feature vector was created, where each feature represented the normalized frequency of a unique n-gram. This was achieved by counting the occurrences of a specific opcode instruction in the opcode file and then dividing this count by the total number of opcode instructions in the file. The acquired data was then inserted into the dataset corresponding to the columns generated by the associated n-gram vocabulary.

\subsection{Dataset Optimization}
To improve the efficiency and computational performance of subsequent classification, the dataset was optimized by removing features with low variance as determined by a threshold based on predefined percentiles. This step ensured that only informative features, which provided meaningful distinctions, contributed to the model training process. The optimized dataset was then stored in \texttt{.csv} and \texttt{.pkl} formats and was ready for use in machine learning models. This optimization helped particularly with the 2-gram dataset, where the number of columns neared sixty thousand, which would have made training take far longer. Below, a table showing the sizes of the original and optimized datasets is provided. For the 1-gram dataset, the 10th percentile variance selection was used, while for the 2-gram dataset, the 80th percentile variance selection was used.

\begin{table}[H]
  \renewcommand{\arraystretch}{1.3}
  \centering
  \caption{Comparison of Dataset Dimensions Before and After Optimization}
  \label{tab:dataset_sizes}
  \begin{tabularx}{\textwidth}{|l|X|X|}
    \hline
    \rowcolor[rgb]{0.85, 0.85, 0.85} 
    \textbf{Dataset} & \textbf{Original Size (Rows × Features)} & \textbf{Optimized Size (Rows × Features)} \\
    \hline
    \rowcolor[rgb]{0.95, 0.95, 0.95} 
    1-Gram & 1930 × 3658 & 1687 × 3658 \\
    \hline
    2-Gram & 59,276 × 3658 & 11,847 × 3685 \\
    \hline
  \end{tabularx}
\end{table}

\subsection{Serialization}
Serialization was extensively used to reduce computation time, trading off increased memory usage for faster performance. When handling large datasets, even seemingly quick computations could accumulate significant overhead, leading to inefficiencies. To avoid redundant recalculations, data was serialized using the pickle format, as it allowed easy serialization and deserialization of Python objects including Pandas DataFrames \cite{temiz2022recording}.

\subsection{Classification}

\subsubsection{Classification Algorithms}
We utilized three types of classification algorithms on the optimized dataset:
\begin{itemize}
    \item \textbf{Support Vector Machine (SVM)}: SVM constructed an optimal hyperplane that maximized the margin between classes. For non-linear separability, kernel functions (such as Radial Basis Function (RBF) or polynomial kernels) were available to transform the feature space \cite{suthaharan2016support}.
    \item \textbf{K-Nearest Neighbors (KNN)}: KNN classified a sample by analyzing the classes of its k-nearest neighbors (using Euclidean distance). Although the provided configuration used \( k = 3 \), the methodology could be adjusted based on empirical validation \cite{peterson2009k}.
    \item \textbf{Decision Tree}: This classifier recursively partitioned the dataset based on feature thresholds, yielding an interpretable tree structure that highlighted the most significant opcode features influencing the decision process \cite{de2013decision}.
\end{itemize}

\subsubsection{Classification Modes}
To evaluate the influence of opcode data and associated metadata, three modes were adopted:

\begin{itemize}
    \item \textbf{Single Mode}: Classification was performed solely on opcode features to predict a target label with metadata (such as APT group, malware name, and malware type) excluded.
    \item \textbf{Multi Mode}: The target label was predicted using opcode features supplemented by metadata from the remaining labels.
    \item \textbf{All Mode}: All label metadata were incorporated into the classification process with the opcode filename serving as the target.
\end{itemize}

For each mode, the classifiers were trained, and predictions were made on the same dataset so that performance metrics could be computed and compared. Since multiple classification algorithms and modes were used, it was determined that the most efficient approach was to create a specialized class to manage the results. The \texttt{ClassifierResult} class encapsulated key performance metrics and stored essential details including the classifier type and classification mode. To ensure both human readability and seamless reusability in Python, the results were serialized in JSON format allowing for both human readability and further computer analysis.

The classification function trained and evaluated a machine learning model using a given dataset, classifier, mode, and target label. It first processed the dataset based on the specified mode (single, multi, or all); for instance, it converted label features to numerical values using ordinal encoding for modes that used metadata as features. Depending on the chosen classifier (SVM, KNN, or Decision Tree), the function trained the model and then made predictions to evaluate performance. The performance metrics included accuracy, recall, precision, F1-score, and confusion matrix. The results, including the trained model and computed metrics, were returned as a \texttt{ClassifierResult} object.

Once the model had been trained and evaluated using the classification function, the results were then visualized in the form of bar charts for accuracy, recall, precision, and F1-score. In terms of the confusion matrix, it was visualized using a heatmap.

\subsection{Hardware Utilized}
Given the large volume of data involved, a high-performance CPU-based system was utilized to minimize computation time.

\begin{table}[H]
\centering
\renewcommand{\arraystretch}{1.4}
\caption{System Hardware Specifications Used for N-Gram Dataset and Training}
\label{tab:hardware-specs}
\begin{tabularx}{\textwidth}{|l|l|X|}
\hline
\rowcolor[rgb]{0.85, 0.85, 0.85}
\textbf{Hardware Type} & \textbf{Model} & \textbf{Specifications} \\
\hline
\rowcolor[rgb]{0.95, 0.95, 0.95}
CPU 1 & Intel Xeon E5-2697 v3 & 14 cores / 28 threads, Base Clock: 2.6GHz, Turbo: 3.6GHz, 35MB Cache \\
\hline
CPU 2 & Intel Xeon E5-2697 v3 & 14 cores / 28 threads, Base Clock: 2.6GHz, Turbo: 3.6GHz, 35MB Cache \\
\hline
\rowcolor[rgb]{0.95, 0.95, 0.95}
Memory & DDR4 RECC & 256GB, 2133MHz \\
\hline
GPU 1 & NVIDIA GeForce RTX 3060 & 12GB GDDR6 VRAM, 3584 CUDA cores, Boost Clock: 1.78GHz \\
\hline
\rowcolor[rgb]{0.95, 0.95, 0.95}
GPU 2 & NVIDIA GeForce RTX 2080 Super & 8GB GDDR6 VRAM, 3072 CUDA cores, Boost Clock: 1.81GHz \\
\hline
Operating System & AlmaLinux & Version 9 (64-bit) \\
\hline
\end{tabularx}
\end{table}

\section{Methodology: Convolutional Neural Network}
\subsection{Preprocessing}
As seen in the results for training using n-grams below, the confusion matrices indicated that traditional machine learning models struggle with one-to-many mapping, particularly when they lack metadata to rely on. In the context of malware analysis, these one-to-many relationships occur when the same malicious executable was associated with multiple APT groups. To examine how this impacted CNN performance, the opcode dataset was preprocessed to remove such overlaps, creating a subset of the dataset with strictly one-to-one mappings to reduce confusion. This filtering process resulted in a 63.75\% reduction in opcode files from our dataset, though many of these were duplicates as the data was organized by APT groups. The reasoning was that carefully curating the dataset to remove ambiguous samples would lead to a more accurate model, even if it meant training on fewer files. Additionally, our initial dataset contained over 3,500 opcode files; therefore, filtering still left a considerable amount of data to work with.

\begin{figure}[H]
    \centering
    \includegraphics[width=\linewidth, keepaspectratio]{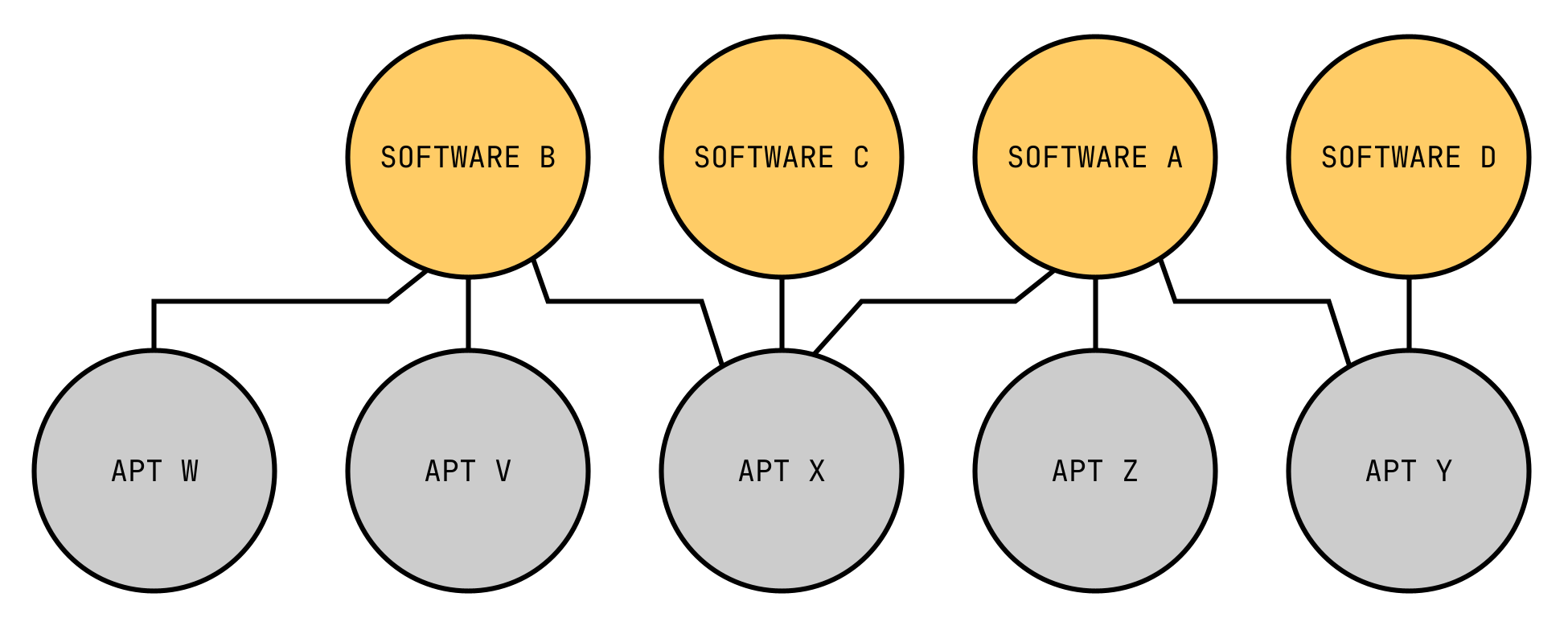}
    \caption{One-to-Many Relationship Between Software and APT Groups}
    \label{fig:one-to-many}
\end{figure}

\begin{figure}[H]
    \centering
    \includegraphics[width=\linewidth, keepaspectratio]{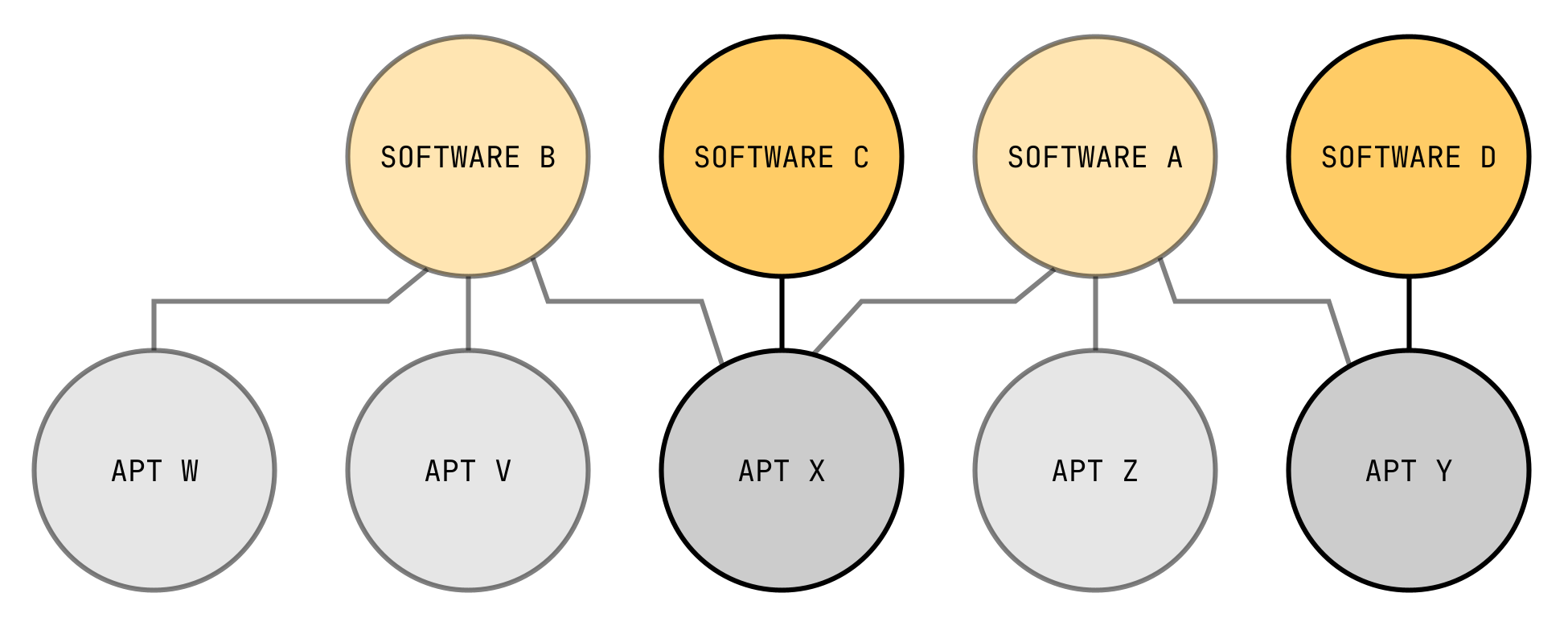}
    \caption{One-to-One Relationship Between Software and APT Groups}
    \label{fig:one-to-one}
\end{figure}

\noindent More specifically, the preprocessing script worked by creating a copy of the original dataset and systematically identifying and removing any duplicate opcode subdirectories pertaining to software belonging to one-to-many relationships. This process ensured that only unique opcodes remained, resulting in a clean, well-structured opcode database that respected a one-to-one mapping. Furthermore, opcode files were refined by stripping whitespace, removing empty lines, and converting opcodes to uppercase to ensure consistency. These processed sequences were then used to generate a structured dataset, where opcodes were grouped based on a selected target label (e.g., APT group, malware name, or malware type).

\subsection{Generating Vocabulary}
Following the methodology used in the paper titled \textit{Deep Android Malware Detection}, the opcodes were encoded as one-hot vectors, where each opcode was represented by a binary vector with all values set to zero except for a single one at the index corresponding to that opcode. This process initially involved identifying all unique opcodes in the dataset and assigning each one an integer index to create a vocabulary. Additionally, labels had to be encoded to be compatible with the model which was done automatically using the label encoding functionality of scikit-learn \cite{deep}.

\begin{figure}[H]
    \centering
    \includegraphics[width=5.5in, keepaspectratio]{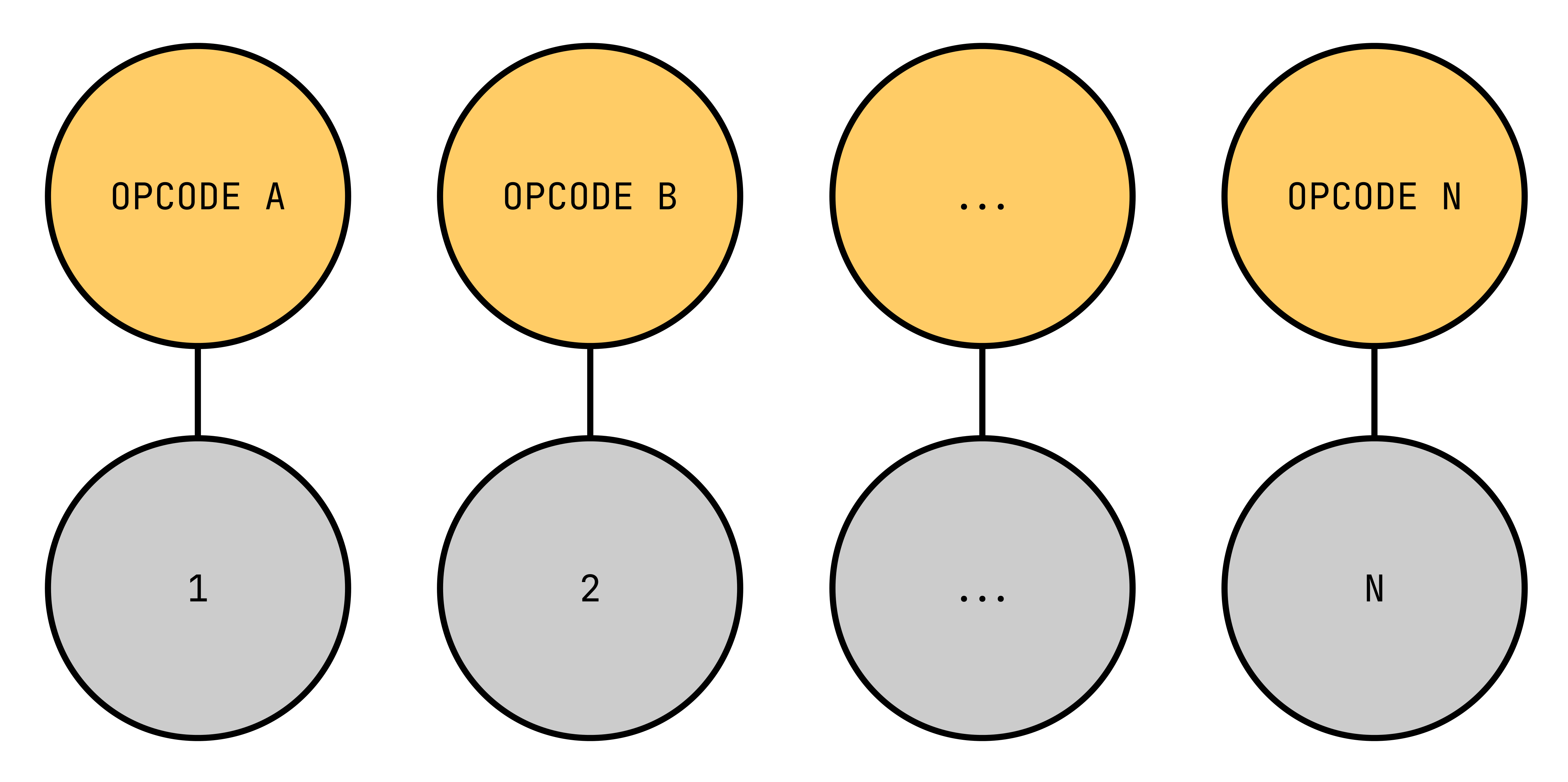}
    \caption{Opcode Vocabulary Mapping}
    \label{fig:vocabulary}
\end{figure}

\subsection{Implementing the CNN Model}
The architecture was inspired by the model presented in the research paper \textit{Deep Android Malware Detection} and was adapted specifically for this case. Unlike the original model, which focused on binary malware detection (benign vs. malicious), this implementation was tailored to classify malware according to specific target labels such as APT group, malware name, or malware type \cite{deep}.

\subsubsection{Embedding Layer}
The embedding layer transformed encoded opcode sequences into dense vector representations making them suitable for processing by convolutional layers.

\subsubsection{Convolutional Layers}
In this implementation, two convolutional layers were applied after the embedding layer enabling the model to hierarchically learn and select important features. The first convolutional layer captured low-level patterns, while the deeper layer extracted higher-level semantic features that were crucial for malware classification. This automated feature extraction process enhanced the deep learning model’s ability to distinguish effectively between different malware categories \cite{gibert2019ml}.

\subsubsection{Pooling Layer}
Pooling is a downsampling technique used in CNNs to reduce the spatial dimensions of feature maps while preserving the most important information. In this implementation, \texttt{MaxPooling1D} was used after each convolutional layer. The pool size of two reduced the sequence length by half, ensuring that only the most important features were retained while minimizing redundancy.

\subsubsection{Flatten Layer}
In this implementation, an additional layer called a flatten layer was added. This served to take the multi-dimensional data generated by the pooling layers and transform it back into one-dimensional data so that the fully connected layer could process the data.

\subsubsection{Fully Connected (MLP) Layer}
These models worked by forming neurons or nodes throughout each layer. The fully connected layer was responsible for creating connections between these layers combining all the learned features so they could be used to classify the input according to the associated labels. The \texttt{Dense} layer enhanced feature representation by introducing non-linearity, while the \texttt{Dropout} layer prevented overfitting by randomly deactivating neurons during training \cite{li2022efficient}.

\subsubsection{Softmax Classification Layer}
The softmax classification layer was the final layer that converted the model’s output into probabilities for each of our labels allowing for multi-class classification.

\begin{figure}[H]
    \centering
    \includegraphics[width=6.2in, keepaspectratio]{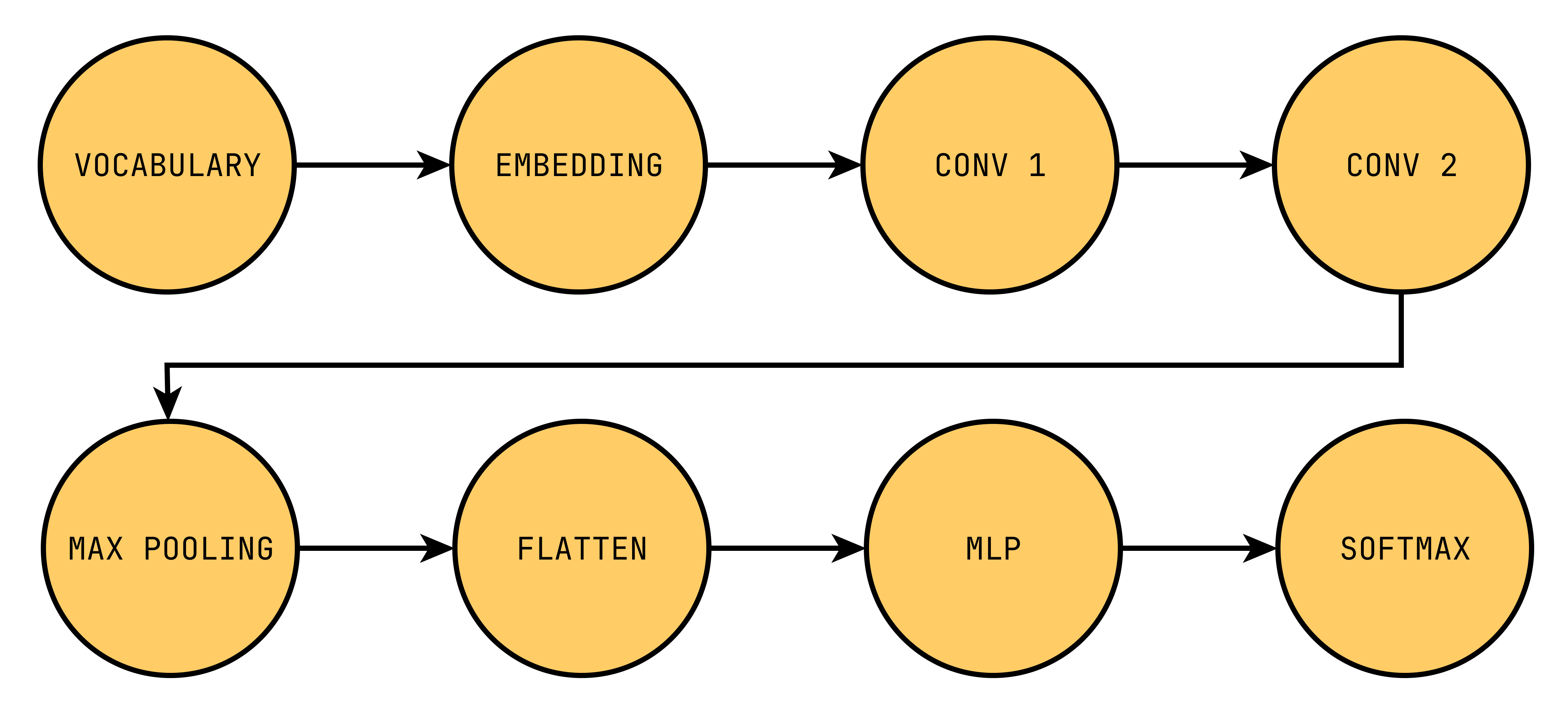}
    \caption{CNN Sequence}
    \label{fig:cnn-sequence}
\end{figure}

\subsection{Training the Model}
In order to train the model, the inputs had to be padded such that their lengths were uniform. To do this, the length of the sequences was selected based on a percentile, as simply choosing the length of the largest sequence would have created much unnecessary padding which could have interfered with the performance of the model. Additionally, malware labels were encoded into a categorical format. The CNN model was trained using the Adam optimizer which adjusted learning rates for faster learning, and categorical cross-entropy loss which was suited for multi-class classification. Together, they helped the model learn patterns in opcode sequences while reducing the chance of overfitting \cite{venkatraman2022opcode}. The model was trained using the following parameters:

\begin{table}[H]
  \centering
  \caption{Parameters Used During CNN Training}
  \renewcommand{\arraystretch}{1.3}
  \begin{tabularx}{\textwidth}{|X|X|}
    \hline
    \rowcolor[rgb]{0.85, 0.85, 0.85} 
    \textbf{Parameter} & \textbf{Value} \\
    \hline
    \rowcolor[rgb]{0.95, 0.95, 0.95} 
    \texttt{k} & 8 \\
    \hline
    \rowcolor[rgb]{0.95, 0.95, 0.95} 
    \texttt{percentile} & 50 \\
    \hline
    \rowcolor[rgb]{0.95, 0.95, 0.95} 
    \texttt{epochs} & 16 \\
    \hline
    \rowcolor[rgb]{0.95, 0.95, 0.95} 
    \texttt{batch\_size} & 32 \\
    \hline
    \rowcolor[rgb]{0.95, 0.95, 0.95} 
    \texttt{validation\_split} & 0.1 \\
    \hline
  \end{tabularx}
  \label{tab:cnn_training_params}
\end{table}

\subsection{Hardware Utilized}
Given the large volume of data to be processed and the model’s capability of leveraging GPU acceleration, the strongest available GPU compute device was selected for use.

\begin{table}[H]
\centering
\renewcommand{\arraystretch}{1.4}
\caption{System Hardware Specifications Used for CNN Training}
\label{tab:hardware-specs}
\begin{tabularx}{\textwidth}{|l|l|X|}
\hline
\rowcolor[rgb]{0.85, 0.85, 0.85}
\textbf{Hardware Type} & \textbf{Model} & \textbf{Specifications} \\
\hline
\rowcolor[rgb]{0.95, 0.95, 0.95}
CPU & AMD Ryzen 9 5950X & 16 cores / 32 threads, Base Clock: 3.4GHz, Boost Clock: up to 4.9GHz \\
\hline
Memory & DDR4 Non-ECC & 128GB, 3600MHz \\
\hline
\rowcolor[rgb]{0.95, 0.95, 0.95}
GPU & NVIDIA RTX 4090 & 24GB GDDR6X VRAM, Boost Clock: up to 2.52GHz, 16384 CUDA cores \\
\hline
Operating System & Fedora & Version 41 (64-bit) \\
\hline
\end{tabularx}
\end{table}

\section{Results: Opcode Extraction}
The script took a total of 18 hours to run was was set to analyze five items at a time. Without parallel processing, it was estimated that this process would have taken at least 90 hours. It should be noted that the large amount of time required was due to limitations in computer hardware as well as the number of malicious executables to be analyzed. Once the script had terminated, a total of 3789 \texttt{.opcode} files organized in the same structure as the original malware dataset remained. This number was lower than the total number of files because non-binary files such as Python scripts were included in the dataset but could not be decompiled by Ghidra, as they are not technically executables but executed with separate applications such as Python.

\section{Results: N-Gram Dataset \& Training}
\subsection{Performance Evaluation}
To enhance understanding of the model's performance both bar charts and heat maps were used. These included all combinations of each mode against each classifier, analyzed for both 1-gram and 2-gram feature extraction techniques. All visualizations can be found in the Appendix of the paper.

\subsection{Preprocessing}
Generating the large datasets from the \texttt{.opcode} files took ten minutes to complete.

\subsection{Training}
Training the models on large datasets was a time-intensive process even after optimizing the datasets. In total, training all combinations of classifiers on the 1-gram dataset took 5.3 minutes, and on the 2-gram dataset it took 12.4 hours.

\subsection{Metrics}
Performance is quantitatively evaluated using:

\begin{itemize}
    \item \textbf{Accuracy}: The proportion of correct predictions (both positive and negative) out of all predictions made \cite{confusionmatrix}.
    \item \textbf{Recall}: The ability of the model to correctly identify all positive instances while minimizing missed positives \cite{confusionmatrix}.
    \item \textbf{Precision}: The ability of the model to only predict positives when they are truly positive while minimizing false positives \cite{confusionmatrix}.
    \item \textbf{F-Measure}: The harmonic mean of precision and recall balancing the trade-off between them for overall performance \cite{confusionmatrix}.
    \item \textbf{Confusion Matrix}: A summary table showing the counts of true positives, true negatives, false positives, and false negatives revealing the types of errors made by the model \cite{confusionmatrix}.
\end{itemize}

This portion of the paper evaluates the performance of three machine learning (ML) classifiers—Support Vector Machine (SVM), K-Nearest Neighbors (KNN), and Decision Tree—for the classification of malware based on opcode sequences extracted from executables associated with APT groups. The classifiers are systematically assessed on their ability to predict three primary targets: malware name, malware type, and APT group attribution, utilizing both 1-gram and 2-gram opcode feature representations to capture the sequential patterns of instruction codes.

\subsection{Classifier Comparison}

\subsubsection{Decision Tree}
The Decision Tree classifier consistently outperformed the other classifier models across both 1-gram and 2-gram datasets demonstrating the highest scores in accuracy, precision, recall, and F-Measure. Notably, the Decision Tree achieved remarkable performance in malware type and malware name classification attaining an accuracy of 99.69\% with an F-Measure of 0.85 for malware type and 97.37\% accuracy with an F-Measure of 0.88 for malware name on the 2-gram dataset using the multi-mode approach. Even when challenged by APT group attribution, it fell only to a moderate F-Measure of around 0.63 still outperforming other models.

\subsubsection{KNN}
KNN demonstrated moderate classification performance, producing acceptable results in both malware name and type classification task; however, it consistently lagged behind the Decision Tree classifier in terms of accuracy and F-Measure.

\subsubsection{SVM}
The SVM classifier demonstrated limited effectiveness across all classification tasks particularly in APT group attribution. The F-Measure for SVM consistently remained below 0.20, suggesting that SVM was not well-suited for capturing the subtle patterns in the opcode data making it less reliable for this task.

\subsection{Mode Impact}

\subsubsection{Single Mode (Opcode Features Only)}
When only opcode-derived features were used, all classifiers struggled to some extent, particularly for APT group attribution. The limited performance in this mode implied that opcode frequencies alone did not provide sufficient context to differentiate between groups especially for threat actors with overlapping characteristics.

\subsubsection{Multi Mode (Opcode Features \& Metadata)}
Incorporating additional metadata significantly improved the classifiers’ performance. This mode showed that adding contextual information such as the malware name and type, could compensate for the limitations of the opcode features. The Decision Tree, in particular, showed remarkable gains, indicating that the combination of opcode data and metadata provided a more complete feature set for accurate classification.

\subsubsection{All Mode (All Labels for File Name Prediction)}
In this mode, where all metadata was included and the file name served as the target, SVM and Decision Tree achieved similar performance (~70\% accuracy). This suggested that when more contextual labels were available, even a weaker classifier like SVM could catch up to a more robust model in terms of overall accuracy; however, KNN remained significantly behind highlighting its vulnerability to the complexity introduced by multiple label dependencies.

\subsection{Raw Performance Data}
\subsubsection{1-Gram Performance Evaluation Table}
A colour-coded table displaying the performance results of the 1-gram models is shown below:
\begin{table}[H]
  \centering
  \caption{1-Gram Classifier Performance}
  \begin{tabularx}{\textwidth}{|X|X|X|X|X|X|X|}
    \hline
    \rowcolor[rgb]{.749, .749, .749}
    \textbf{Classifier} & \textbf{Mode} & \textbf{Target} & \textbf{Accuracy} & \textbf{Recall} & \textbf{Precision} & \textbf{F-Measure} \\
    \hline
    \rowcolor[rgb]{.949, .949, .949} SVM & Single & Group & \cellcolor[rgb]{.98, .616, .459}0.2797 & \cellcolor[rgb]{.973, .424, .42}0.0741 & \cellcolor[rgb]{.973, .424, .42}0.0728 & \cellcolor[rgb]{.973, .412, .42}0.0587 \\
    \hline
    \rowcolor[rgb]{.949, .949, .949} SVM & Single & Name & \cellcolor[rgb]{.996, .851, .502}0.5295 & \cellcolor[rgb]{.973, .416, .42}0.0661 & \cellcolor[rgb]{.973, .451, .427}0.1039 & \cellcolor[rgb]{.973, .42, .42}0.0686 \\
    \hline
    \rowcolor[rgb]{.949, .949, .949} SVM & Single & Type & \cellcolor[rgb]{.784, .859, .506}0.7438 & \cellcolor[rgb]{.976, .482, .431}0.1377 & \cellcolor[rgb]{.98, .58, .451}0.2419 & \cellcolor[rgb]{.976, .49, .431}0.143 \\
    \hline
    \rowcolor[rgb]{.949, .949, .949} KNN & Single & Group & \cellcolor[rgb]{.988, .753, .482}0.4248 & \cellcolor[rgb]{.992, .804, .494}0.4789 & \cellcolor[rgb]{.992, .831, .498}0.5098 & \cellcolor[rgb]{.988, .733, .478}0.4028 \\
    \hline
    \rowcolor[rgb]{.949, .949, .949} KNN & Single & Name & \cellcolor[rgb]{.608, .808, .498}0.8576 & \cellcolor[rgb]{.984, .702, .475}0.3693 & \cellcolor[rgb]{.988, .718, .478}0.3885 & \cellcolor[rgb]{.984, .694, .471}0.3612 \\
    \hline
    \rowcolor[rgb]{.949, .949, .949} KNN & Single & Type & \cellcolor[rgb]{.451, .765, .486}0.9584 & \cellcolor[rgb]{.976, .918, .518}0.6202 & \cellcolor[rgb]{.949, .91, .518}0.6376 & \cellcolor[rgb]{.984, .918, .518}0.6164 \\
    \hline
    \rowcolor[rgb]{.949, .949, .949} Decision Tree & Single & Group & \cellcolor[rgb]{.992, .839, .502}0.5197 & \cellcolor[rgb]{.933, .902, .514}0.6482 & \cellcolor[rgb]{.824, .871, .51}0.7194 & \cellcolor[rgb]{.996, .914, .514}0.5972 \\
    \hline
    \rowcolor[rgb]{.949, .949, .949} Decision Tree & Single & Name & \cellcolor[rgb]{.459, .765, .486}0.9535 & \cellcolor[rgb]{.553, .792, .494}0.8925 & \cellcolor[rgb]{.655, .824, .498}0.8258 & \cellcolor[rgb]{.631, .816, .498}0.8415 \\
    \hline
    \rowcolor[rgb]{.949, .949, .949} Decision Tree & Single & Type & \cellcolor[rgb]{.4, .749, .486}0.9896 & \cellcolor[rgb]{.537, .788, .494}0.9022 & \cellcolor[rgb]{.647, .82, .498}0.8318 & \cellcolor[rgb]{.608, .812, .498}0.8563 \\
    \hline
    \rowcolor[rgb]{.949, .949, .949} SVM & Multi & Group & \cellcolor[rgb]{.988, .706, .475}0.3751 & \cellcolor[rgb]{.976, .486, .431}0.1405 & \cellcolor[rgb]{.973, .443, .424}0.0964 & \cellcolor[rgb]{.973, .459, .427}0.1099 \\
    \hline
    \rowcolor[rgb]{.949, .949, .949} SVM & Multi & Name & \cellcolor[rgb]{.992, .922, .518}0.6107 & \cellcolor[rgb]{.973, .424, .42}0.0716 & \cellcolor[rgb]{.973, .416, .42}0.0631 & \cellcolor[rgb]{.973, .412, .42}0.0614 \\
    \hline
    \rowcolor[rgb]{.949, .949, .949} SVM & Multi & Type & \cellcolor[rgb]{.855, .882, .51}0.6996 & \cellcolor[rgb]{.973, .451, .424}0.1007 & \cellcolor[rgb]{.976, .549, .443}0.2092 & \cellcolor[rgb]{.973, .443, .424}0.0936 \\
    \hline
    \rowcolor[rgb]{.949, .949, .949} KNN & Multi & Group & \cellcolor[rgb]{.992, .831, .498}0.5082 & \cellcolor[rgb]{.992, .831, .498}0.508 & \cellcolor[rgb]{.996, .878, .51}0.5611 & \cellcolor[rgb]{.988, .773, .486}0.4458 \\
    \hline
    \rowcolor[rgb]{.949, .949, .949} KNN & Multi & Name & \cellcolor[rgb]{.592, .804, .494}0.8666 & \cellcolor[rgb]{.988, .729, .478}0.4002 & \cellcolor[rgb]{.988, .737, .478}0.4072 & \cellcolor[rgb]{.988, .722, .478}0.3918 \\
    \hline
    \rowcolor[rgb]{.949, .949, .949} KNN & Multi & Type & \cellcolor[rgb]{.424, .757, .486}0.9751 & \cellcolor[rgb]{.875, .886, .514}0.6852 & \cellcolor[rgb]{.769, .855, .506}0.7537 & \cellcolor[rgb]{.855, .882, .51}0.6977 \\
    \hline
    \rowcolor[rgb]{.949, .949, .949} Decision Tree & Multi & Group & \cellcolor[rgb]{.996, .851, .502}0.5287 & \cellcolor[rgb]{.863, .882, .51}0.6925 & \cellcolor[rgb]{.757, .855, .506}0.761 & \cellcolor[rgb]{.953, .91, .518}0.6359 \\
    \hline
    \rowcolor[rgb]{.949, .949, .949} Decision Tree & Multi & Name & \cellcolor[rgb]{.427, .757, .486}0.9738 & \cellcolor[rgb]{.475, .773, .49}0.9435 & \cellcolor[rgb]{.584, .804, .494}0.8718 & \cellcolor[rgb]{.557, .796, .494}0.8898 \\
    \hline
    \rowcolor[rgb]{.949, .949, .949} Decision Tree & Multi & Type & \cellcolor[rgb]{.388, .745, .482}0.997 & \cellcolor[rgb]{.529, .788, .494}0.9066 & \cellcolor[rgb]{.647, .82, .498}0.833 & \cellcolor[rgb]{.604, .808, .498}0.8592 \\
    \hline
    \rowcolor[rgb]{.949, .949, .949} SVM & All & File Name & \cellcolor[rgb]{.839, .875, .51}0.7094 & \cellcolor[rgb]{.839, .875, .51}0.7094 & \cellcolor[rgb]{.914, .898, .514}0.6603 & \cellcolor[rgb]{.894, .894, .514}0.6728 \\
    \hline
    \rowcolor[rgb]{.949, .949, .949} KNN & All & File Name & \cellcolor[rgb]{.98, .596, .455}0.2586 & \cellcolor[rgb]{.98, .596, .455}0.2586 & \cellcolor[rgb]{.973, .467, .427}0.1208 & \cellcolor[rgb]{.976, .498, .435}0.1525 \\
    \hline
    \rowcolor[rgb]{.949, .949, .949} Decision Tree & All & File Name & \cellcolor[rgb]{.839, .875, .51}0.7094 & \cellcolor[rgb]{.839, .875, .51}0.7094 & \cellcolor[rgb]{.914, .898, .514}0.6603 & \cellcolor[rgb]{.894, .894, .514}0.6728 \\
    \hline
  \end{tabularx}
  \label{tab:addlabel}
\end{table}

\subsubsection{2-Gram Performance Evaluation Table}
A colour-coded table displaying the performance results of the 2-gram models is shown below:

\begin{table}[H]
  \centering
  \caption{2-Gram Classifier Performance}
  \begin{tabularx}{\textwidth}{|X|X|X|X|X|X|X|}
    \hline
    \rowcolor[rgb]{.749, .749, .749}
    \textbf{Classifier} & \textbf{Mode} & \textbf{Target} & \textbf{Accuracy} & \textbf{Recall} & \textbf{Precision} & \textbf{F-Measure} \\
    \hline
    \rowcolor[rgb]{.949, .949, .949} SVM & Single & Group & \cellcolor[rgb]{.98, .627, .459}0.2914 & \cellcolor[rgb]{.973, .427, .42}0.0807 & \cellcolor[rgb]{.976, .502, .435}0.1599 & \cellcolor[rgb]{.973, .416, .42}0.0684 \\
    \hline
    \rowcolor[rgb]{.949, .949, .949} SVM & Single & Name & \cellcolor[rgb]{.996, .89, .51}0.573 & \cellcolor[rgb]{.973, .424, .42}0.0757 & \cellcolor[rgb]{.973, .467, .427}0.1217 & \cellcolor[rgb]{.973, .427, .42}0.0791 \\
    \hline
    \rowcolor[rgb]{.949, .949, .949} SVM & Single & Type & \cellcolor[rgb]{.702, .835, .502}0.7963 & \cellcolor[rgb]{.976, .51, .435}0.1686 & \cellcolor[rgb]{.984, .667, .467}0.3353 & \cellcolor[rgb]{.976, .529, .439}0.1907 \\
    \hline
    \rowcolor[rgb]{.949, .949, .949} KNN & Single & Group & \cellcolor[rgb]{.988, .753, .482}0.4276 & \cellcolor[rgb]{.992, .8, .494}0.4755 & \cellcolor[rgb]{.996, .863, .506}0.5415 & \cellcolor[rgb]{.988, .737, .482}0.4118 \\
    \hline
    \rowcolor[rgb]{.949, .949, .949} KNN & Single & Name & \cellcolor[rgb]{.604, .808, .498}0.8609 & \cellcolor[rgb]{.988, .729, .478}0.4023 & \cellcolor[rgb]{.988, .737, .478}0.4084 & \cellcolor[rgb]{.988, .718, .478}0.3907 \\
    \hline
    \rowcolor[rgb]{.949, .949, .949} KNN & Single & Type & \cellcolor[rgb]{.443, .765, .486}0.962 & \cellcolor[rgb]{.953, .91, .518}0.6353 & \cellcolor[rgb]{.902, .894, .514}0.6675 & \cellcolor[rgb]{.949, .91, .518}0.639 \\
    \hline
    \rowcolor[rgb]{.949, .949, .949} Decision Tree & Single & Group & \cellcolor[rgb]{.992, .839, .502}0.5197 & \cellcolor[rgb]{.933, .902, .514}0.6482 & \cellcolor[rgb]{.824, .871, .51}0.7194 & \cellcolor[rgb]{.996, .914, .514}0.5972 \\
    \hline
    \rowcolor[rgb]{.949, .949, .949} Decision Tree & Single & Name & \cellcolor[rgb]{.459, .765, .486}0.9535 & \cellcolor[rgb]{.553, .792, .494}0.8925 & \cellcolor[rgb]{.655, .824, .498}0.8258 & \cellcolor[rgb]{.631, .816, .498}0.8415 \\
    \hline
    \rowcolor[rgb]{.949, .949, .949} Decision Tree & Single & Type & \cellcolor[rgb]{.4, .749, .486}0.9896 & \cellcolor[rgb]{.537, .788, .494}0.9022 & \cellcolor[rgb]{.647, .82, .498}0.8318 & \cellcolor[rgb]{.608, .812, .498}0.8563 \\
    \hline
    \rowcolor[rgb]{.949, .949, .949} SVM & Multi & Group & \cellcolor[rgb]{.988, .706, .475}0.3751 & \cellcolor[rgb]{.976, .482, .431}0.1405 & \cellcolor[rgb]{.973, .443, .424}0.0964 & \cellcolor[rgb]{.973, .455, .427}0.1099 \\
    \hline
    \rowcolor[rgb]{.949, .949, .949} SVM & Multi & Name & \cellcolor[rgb]{.992, .922, .518}0.611 & \cellcolor[rgb]{.973, .42, .42}0.0716 & \cellcolor[rgb]{.973, .412, .42}0.0633 & \cellcolor[rgb]{.973, .412, .42}0.0615 \\
    \hline
    \rowcolor[rgb]{.949, .949, .949} SVM & Multi & Type & \cellcolor[rgb]{.855, .882, .51}0.6996 & \cellcolor[rgb]{.973, .447, .424}0.1007 & \cellcolor[rgb]{.976, .549, .443}0.2092 & \cellcolor[rgb]{.973, .439, .424}0.0936 \\
    \hline
    \rowcolor[rgb]{.949, .949, .949} KNN & Multi & Group & \cellcolor[rgb]{.992, .831, .498}0.5085 & \cellcolor[rgb]{.992, .827, .498}0.5059 & \cellcolor[rgb]{.996, .859, .506}0.5399 & \cellcolor[rgb]{.988, .761, .486}0.436 \\
    \hline
    \rowcolor[rgb]{.949, .949, .949} KNN & Multi & Name & \cellcolor[rgb]{.584, .804, .494}0.8723 & \cellcolor[rgb]{.988, .753, .482}0.4281 & \cellcolor[rgb]{.988, .765, .486}0.4399 & \cellcolor[rgb]{.988, .745, .482}0.4192 \\
    \hline
    \rowcolor[rgb]{.949, .949, .949} KNN & Multi & Type & \cellcolor[rgb]{.427, .757, .486}0.974 & \cellcolor[rgb]{.859, .882, .51}0.6956 & \cellcolor[rgb]{.784, .859, .506}0.7441 & \cellcolor[rgb]{.863, .882, .51}0.6939 \\
    \hline
    \rowcolor[rgb]{.949, .949, .949} Decision Tree & Multi & Group & \cellcolor[rgb]{.996, .847, .502}0.5287 & \cellcolor[rgb]{.863, .882, .51}0.6925 & \cellcolor[rgb]{.757, .855, .506}0.761 & \cellcolor[rgb]{.953, .91, .518}0.6359 \\
    \hline
    \rowcolor[rgb]{.949, .949, .949} Decision Tree & Multi & Name & \cellcolor[rgb]{.427, .757, .486}0.9738 & \cellcolor[rgb]{.475, .773, .49}0.9435 & \cellcolor[rgb]{.584, .804, .494}0.8718 & \cellcolor[rgb]{.557, .796, .494}0.8898 \\
    \hline
    \rowcolor[rgb]{.949, .949, .949} Decision Tree & Multi & Type & \cellcolor[rgb]{.388, .745, .482}0.997 & \cellcolor[rgb]{.529, .788, .494}0.9066 & \cellcolor[rgb]{.647, .82, .498}0.833 & \cellcolor[rgb]{.604, .808, .498}0.8592 \\
    \hline
    \rowcolor[rgb]{.949, .949, .949} SVM & All & File Name & \cellcolor[rgb]{.839, .875, .51}0.7094 & \cellcolor[rgb]{.839, .875, .51}0.7094 & \cellcolor[rgb]{.914, .898, .514}0.6603 & \cellcolor[rgb]{.894, .894, .514}0.6728 \\
    \hline
    \rowcolor[rgb]{.949, .949, .949} KNN & All & File Name & \cellcolor[rgb]{.98, .592, .451}0.2542 & \cellcolor[rgb]{.98, .592, .451}0.2542 & \cellcolor[rgb]{.973, .463, .427}0.118 & \cellcolor[rgb]{.976, .494, .435}0.1492 \\
    \hline
    \rowcolor[rgb]{.949, .949, .949} Decision Tree & All & File Name & \cellcolor[rgb]{.839, .875, .51}0.7094 & \cellcolor[rgb]{.839, .875, .51}0.7094 & \cellcolor[rgb]{.914, .898, .514}0.6603 & \cellcolor[rgb]{.894, .894, .514}0.6728 \\
    \hline
  \end{tabularx}
  \label{tab:addlabel}
\end{table}

\subsection{Performance Summary}
From this, it was understood that Decision Trees excelled in opcode-based malware classification especially when enriched with metadata, while SVM underperformed and KNN showed moderate results. Nevertheless, accurately attributing APT groups remained challenging, indicating that additional contextual data was needed.

\section{Results: Convolutional Neural Network}
Experiments evaluated the CNN model on two distinct datasets: one with a one-to-one mapping (cleaned opcode sequences) and one with a one-to-many mapping (raw opcode sequences). For each dataset, the model was assessed on three target labels—APT group, malware name, and malware type—using standard performance metrics (accuracy, precision, recall, and F1-score).

Compared to traditional classifiers, the implementation of a deep learning model significantly accelerated the process of feature extraction and training by eliminating the need for manual feature selection. By leveraging GPU acceleration, training time was orders of magnitude faster enabling efficient model construction and optimization \cite{li2022efficient}.

\subsection{Preprocessing}
Cleaning the dataset to create a one-to-one mapping took the system 1.53 minutes. This step drastically reduced confusion in the data particularly for the APT group target.

\subsection{Training}
The training process was highly resource-intensive with the workstation drawing approximately 600W of power, fully utilizing the GPU’s 24GB of VRAM, and consuming 50GB of system memory. The substantial resource demand enabled the model to complete training in 3.65 minutes on all targets. This performance significantly outperformed the training times reported in \textit{Deep Android Malware Detection}, highlighting the advantages of modern hardware in efficiently training complex models.

\subsection{Dataset Comparison}
As expected, the one-to-one dataset significantly outperformed the one-to-many dataset in results for the group target, with the one-to-many relationship achieving only 35\% accuracy, while the one-to-one relationship achieved 92\% accuracy. This improvement was attributed to the removal of ambiguity in the data allowing the CNN to accurately categorize samples into APT groups. Meanwhile, name and type classification showed relatively stable performance across both datasets achieving 84.64\% and 89.80\% in the one-to-one dataset compared to 85.69\% and 96.50\% in the one-to-many dataset. These labels typically did not involve overlapping mappings and were therefore less susceptible to the confusion that affected the group target.

\subsubsection{One-To-Many Dataset}

\begin{table}[H]
  \centering
  \caption{CNN Performance With One-to-Many Dataset}
  \begin{tabularx}{\textwidth}{|X|X|X|X|X|X|X|}
    \hline
    \rowcolor[rgb]{.8, .8, .8}
    \textbf{Classifier} & \textbf{Mode} & \textbf{Target} & \textbf{Accuracy} & \textbf{Recall} & \textbf{Precision} & \textbf{F-Measure} \\
    \hline
    \rowcolor[rgb]{.949, .949, .949}
    CNN & Single & Group & \cellcolor[rgb]{.976, .51, .435}0.35 & \cellcolor[rgb]{.973, .42, .42}0.2428 & \cellcolor[rgb]{.976, .51, .435}0.35 & \cellcolor[rgb]{.973, .412, .42}0.2335 \\
    \hline
    \rowcolor[rgb]{.949, .949, .949}
    CNN & Single & Name & \cellcolor[rgb]{.91, .898, .514}0.8569 & \cellcolor[rgb]{.996, .886, .51}0.7985 & \cellcolor[rgb]{.91, .898, .514}0.8569 & \cellcolor[rgb]{.996, .902, .514}0.8177 \\
    \hline
    \rowcolor[rgb]{.949, .949, .949}
    CNN & Single & Type & \cellcolor[rgb]{.388, .745, .482}0.965 & \cellcolor[rgb]{.443, .765, .486}0.9536 & \cellcolor[rgb]{.388, .745, .482}0.965 & \cellcolor[rgb]{.42, .757, .486}0.9588 \\
    \hline
  \end{tabularx}
  \label{tab:addlabel}
\end{table}

\begin{figure}[H]
    \centering
    \includegraphics[width=5.5in, keepaspectratio]{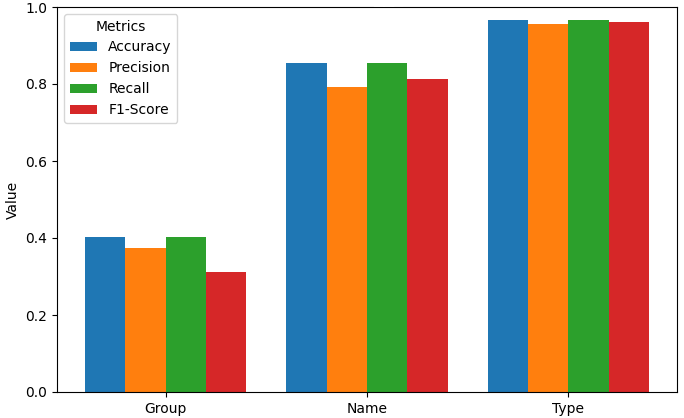}
    \caption{Metric Comparison With One-To-Many Dataset}
    \label{fig:comparision-base}
\end{figure}

\subsubsection{One-To-One Dataset}

\begin{table}[H]
  \centering
  \caption{CNN Performance With One-to-One Dataset}
  \begin{tabularx}{\textwidth}{|X|X|X|X|X|X|X|}
    \hline
    \rowcolor[rgb]{.8, .8, .8}
    \textbf{Classifier} & \textbf{Mode} & \textbf{Target} & \textbf{Accuracy} & \textbf{Recall} & \textbf{Precision} & \textbf{F-Measure} \\
    \hline
    \rowcolor[rgb]{.949, .949, .949}
    CNN & Single & Group & \cellcolor[rgb]{.388, .745, .482}0.9215 & \cellcolor[rgb]{.392, .749, .486}0.9214 & \cellcolor[rgb]{.388, .745, .482}0.9215 & \cellcolor[rgb]{.486, .776, .49}0.9162 \\
    \hline
    \rowcolor[rgb]{.949, .949, .949}
    CNN & Single & Name & \cellcolor[rgb]{.98, .612, .455}0.8464 & \cellcolor[rgb]{.973, .478, .431}0.828 & \cellcolor[rgb]{.98, .612, .455}0.8464 & \cellcolor[rgb]{.973, .412, .42}0.8188 \\
    \hline
    \rowcolor[rgb]{.949, .949, .949}
    CNN & Single & Type & \cellcolor[rgb]{.82, .871, .51}0.898 & \cellcolor[rgb]{.992, .847, .502}0.8779 & \cellcolor[rgb]{.82, .871, .51}0.898 & \cellcolor[rgb]{.992, .824, .498}0.8748 \\
    \hline
  \end{tabularx}
  \label{tab:addlabel}
\end{table}

\begin{figure}[H]
    \centering
    \includegraphics[width=5.5in, keepaspectratio]{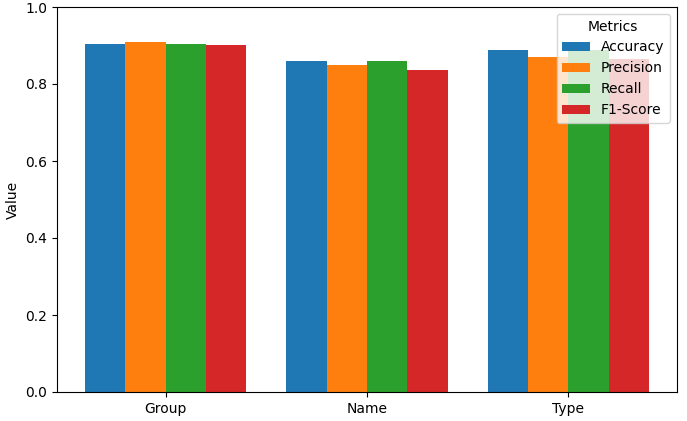}
    \caption{Metric Comparison With One-To-One Dataset}
    \label{fig:comparision-cleaned}
\end{figure}

\section{Discussion: Opcode Extraction}
This paper utilized open-source decompiler technologies and leveraged parallel computing to efficiently and effectively extract opcodes from malicious executables. The results demonstrated that, if scripts are designed to utilize and effectively manage hardware resources then opcode extraction of large datasets is achievable within a reasonable time frame. This opcode extraction provides a strong foundation for further malware analysis using classification algorithms and advanced machine learning techniques.

\section{Discussion: N-Gram Dataset \& Training}
The results from the experiment demonstrated that the Decision Tree classifier consistently outperformed both SVM and KNN for malware classification using opcode features. In particular, when using 2-gram features along with metadata in a specified target (multi-mode) approach, the Decision Tree model achieved exceptional accuracy (up to 99.69\% for malware type and 97.37\% for malware name) and strong F-measures. While KNN produced moderate results, SVM struggled across all tasks with F-Measure values below 0.20 especially for APT group attribution.

Integrating metadata with opcode features significantly improved classifier performance compared to using opcode features alone; however, classifying APT groups remained a challenge, indicating that opcode patterns alone may not fully capture the nuances required for accurate threat actor identification. This difficulty stemmed primarily from the fact that multiple APT groups often used the same software which was included in the dataset to ensure realism. Consequently, it was expected that the models would struggle to accurately predict the specific APT group, as there was not a one-to-one mapping; several APT groups could be associated with the same set of malware. This was further demonstrated by inspecting the confusion matrix heatmaps in the Appendix associated with the group target label.

\section{Discussion: Convolutional Neutral Networks}
The results of the experiment showed that the CNN drastically outperformed traditional machine learning classifiers such as SVM, KNN, and Decision Tree when considering pure opcode analysis without the use of metadata.

\subsection{Metric Comparison}
When attempting to classify malware using traditional methods solely based on pure opcodes without the help of metadata, these models struggled considerably. This is not to say that the models themselves are flawed; rather, they were provided with lower-quality data compared to the CNN. First, the data fed to such models was not cleaned to the extent that the CNN datasets were, and numerous one-to-many relationships were present. In addition, the datasets were limited to 1-gram and 2-gram structures restricting each model’s view to very short opcode windows. Extending n-grams beyond 2-gram was not feasible due to the immense resource overhead it would require, especially since moving from 1-gram to 2-gram alone already increased the feature space into the tens of thousands of features \cite{gibert2019ml}.

Regarding CNN performance, it vastly outperformed the traditional models across multiple targets. For APT group classification specifically, the CNN achieved a 77.65\% higher accuracy compared to the best Decision Tree on uncleaned datasets (from ~52\% to 92\%). In terms of precision and recall, the CNN consistently scored above 0.90 on one-to-one data for this target, whereas Decision Tree scores ranged between 0.64 and 0.72. For other labels such as name and type, the CNN similarly maintained a higher F1-score, in some cases surpassing 0.95, while the best traditional models hovered around the mid-0.80 range. These improvements highlight the importance of extended sequence analysis for malware detection, as CNNs are not limited to small, rigid n-gram windows.

In real-world scenarios, metadata such as file origin or known associations may not be accessible. Traditional models have shown that without metadata, their accuracy can drop to near chance levels for complex tasks like APT group attribution. By contrast, the CNN was able to learn from purely sequential opcode data showcasing its robustness and adaptability. This underscores the significance of leveraging models capable of capturing deeper, long-range dependencies—particularly in scenarios where additional context or metadata is unavailable \cite{li2022augmentation}.

\subsection{Computational Comparison}
Leveraging GPU acceleration for training the CNN models resulted in a substantial reduction in computational time compared to traditional classifiers. As detailed in the results section, the CNN completed training in mere minutes, although it consumed significant amounts of power, while the traditional models required a total of 12.4 hours to train. This contrast in training times was simply incomparable, especially considering that the CNN demonstrated superior performance in pure opcode analysis further emphasizing the efficiency and effectiveness of deep learning approaches over conventional methods.

\subsection{Concluding Remarks}
This paper, along with the broader scope of the project, demonstrated that real-world malware collected from the internet could be effectively classified using only raw opcodes. The implementation of such machine learning methods holds significant potential for practical application in malware detection systems; however, training these models remains computationally demanding often limiting their accessibility to researchers and large organizations. Continued research is therefore essential to develop more efficient algorithms, and reduce the computational cost of training and deploying models in real-world environments.

Future work will explore other deep learning architectures particularly recurrent neural networks (RNNs) and transformers. While RNNs (such as LSTMs or GRUs) can learn sequence patterns with fewer resources, transformers excel at capturing long-range dependencies in a highly parallelizable manner. Moreover, future efforts will focus on incorporating multi-label classification techniques such as using a sigmoid output layer with threshold tuning or leveraging classifier chains to better address real-world scenarios where malware may overlap across multiple threat actor categories. By experimenting with these architectures and broader labeling strategies, the goal is to refine the balance between accuracy and computational overhead paving the way for broader deployment of advanced malware detection solutions.

\newpage
\bibliographystyle{IEEEtran}
\bibliography{references}

\newpage

\appendix
\section{Appendix: 1-Gram Performance Visualizations}
\subsection*{Classifier Performance Visualizations}

\begin{figure}[H]
  \centering
  \begin{minipage}[t]{0.48\textwidth}
    \centering
    \includegraphics[width=\linewidth]{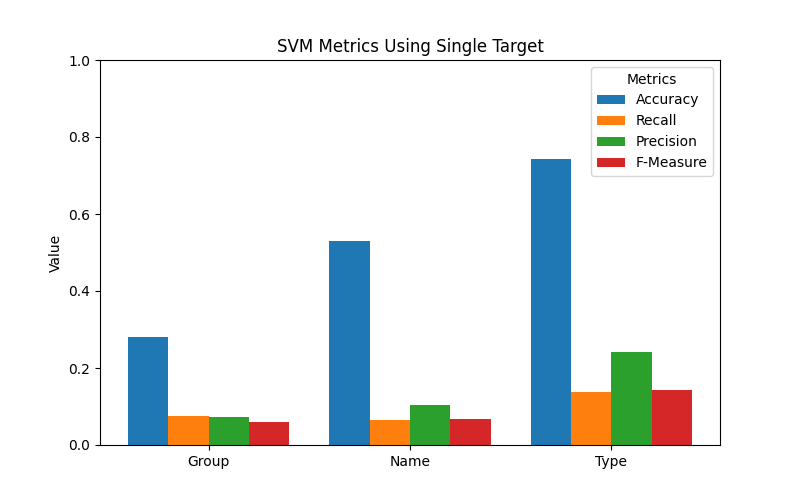}
    \captionof{figure}{SVM – Single Target}
  \end{minipage}\hfill
  \begin{minipage}[t]{0.48\textwidth}
    \centering
    \includegraphics[width=\linewidth]{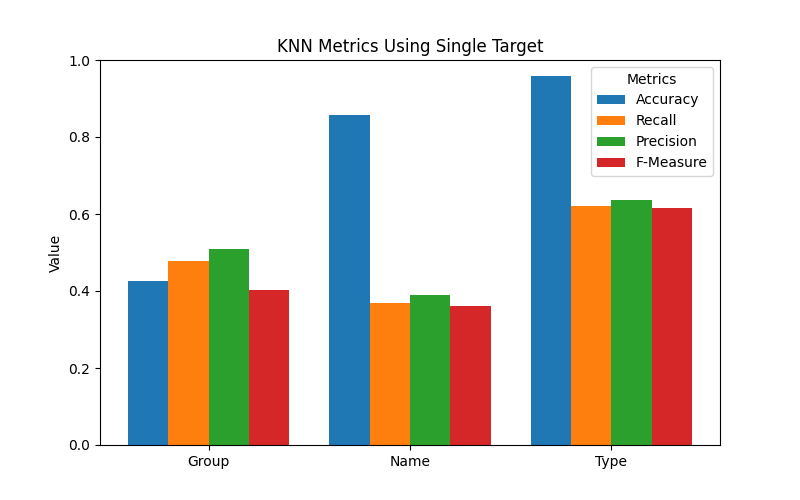}
    \captionof{figure}{KNN – Single Target}
  \end{minipage}
\end{figure}

\begin{figure}[H]
  \centering
  \begin{minipage}[t]{0.48\textwidth}
    \centering
    \includegraphics[width=\linewidth]{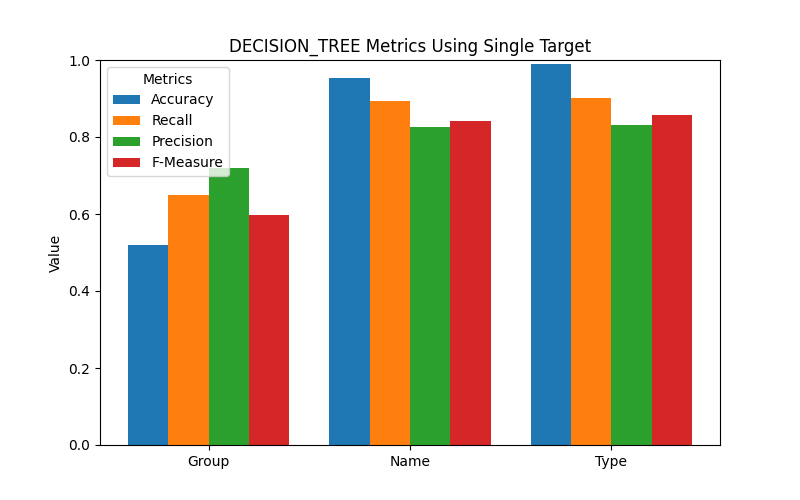}
    \captionof{figure}{Decision Tree – Single Target}
  \end{minipage}\hfill
  \begin{minipage}[t]{0.48\textwidth}
    \centering
    \includegraphics[width=\linewidth]{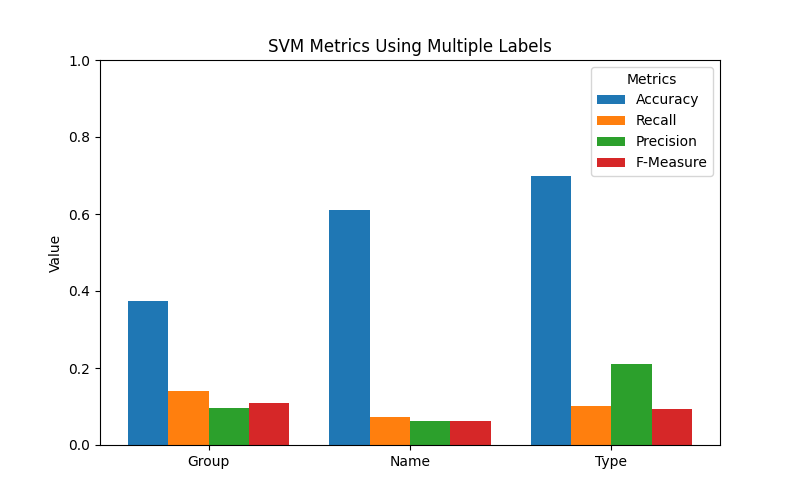}
    \captionof{figure}{SVM – Multi-Label}
  \end{minipage}
\end{figure}

\begin{figure}[H]
  \centering
  \begin{minipage}[t]{0.48\textwidth}
    \centering
    \includegraphics[width=\linewidth]{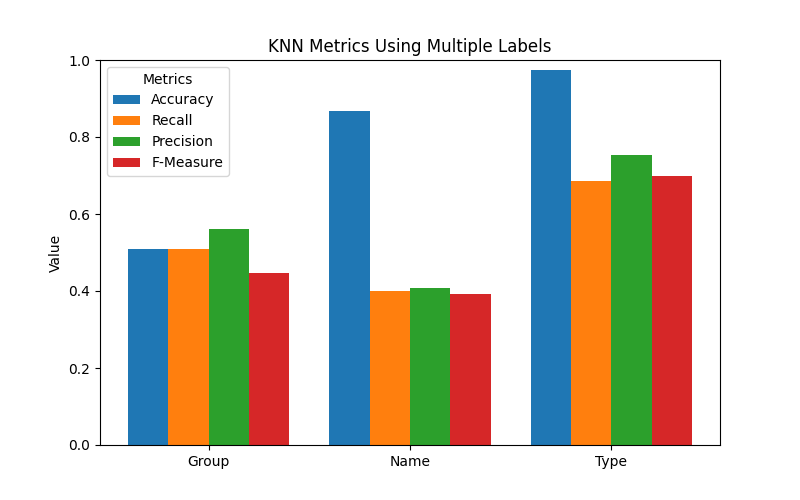}
    \captionof{figure}{KNN – Multi-Label}
  \end{minipage}\hfill
  \begin{minipage}[t]{0.48\textwidth}
    \centering
    \includegraphics[width=\linewidth]{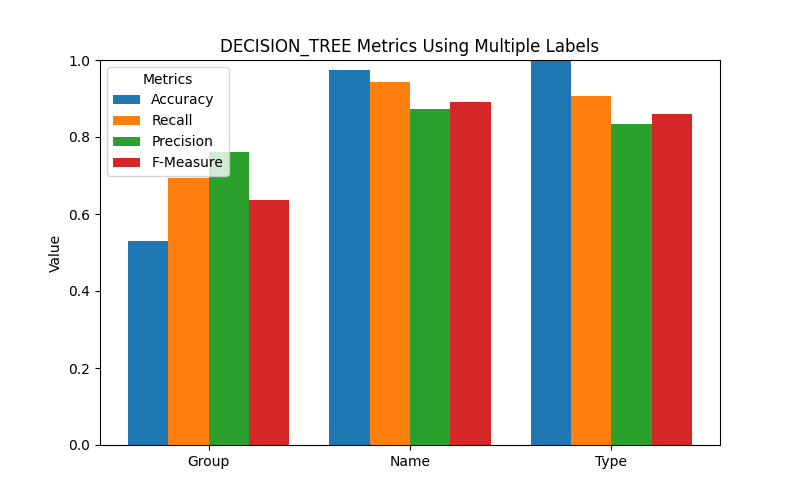}
    \captionof{figure}{Decision Tree – Multi-Label}
  \end{minipage}
\end{figure}

\begin{figure}[H]
  \centering
  \begin{minipage}[t]{0.48\textwidth}
    \centering
    \includegraphics[width=\linewidth]{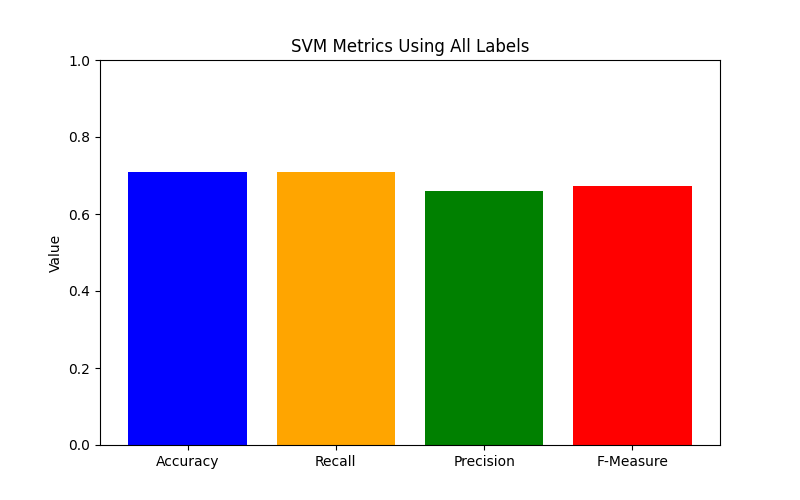}
    \captionof{figure}{SVM – All Labels}
  \end{minipage}\hfill
  \begin{minipage}[t]{0.48\textwidth}
    \centering
    \includegraphics[width=\linewidth]{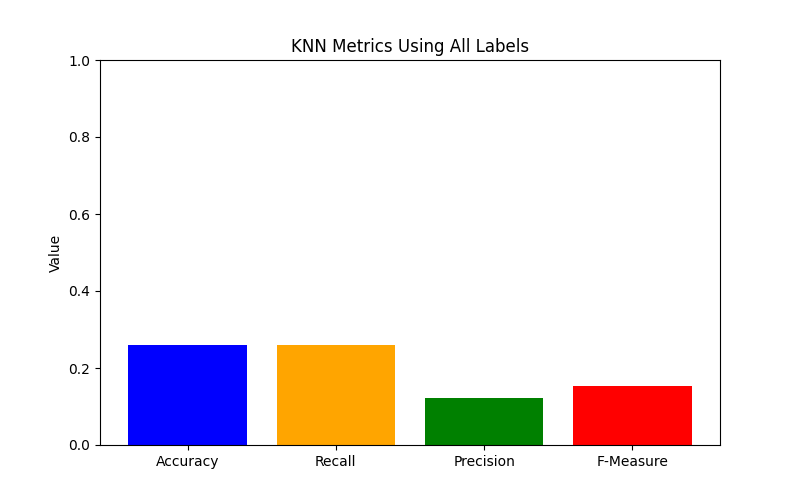}
    \captionof{figure}{KNN – All Labels}
  \end{minipage}
\end{figure}

\begin{figure}[H]
  \centering
  \includegraphics[width=0.48\textwidth]{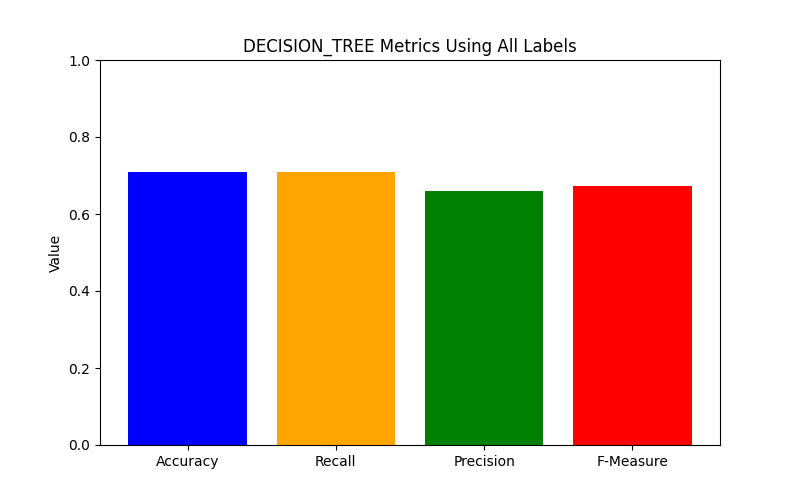}
  \caption{Decision Tree – All Labels}
\end{figure}

\FloatBarrier

\subsection*{Confusion Matrices}

\begin{figure}[H]
  \centering
  \begin{minipage}[t]{0.32\textwidth}\centering
    \includegraphics[width=\linewidth]{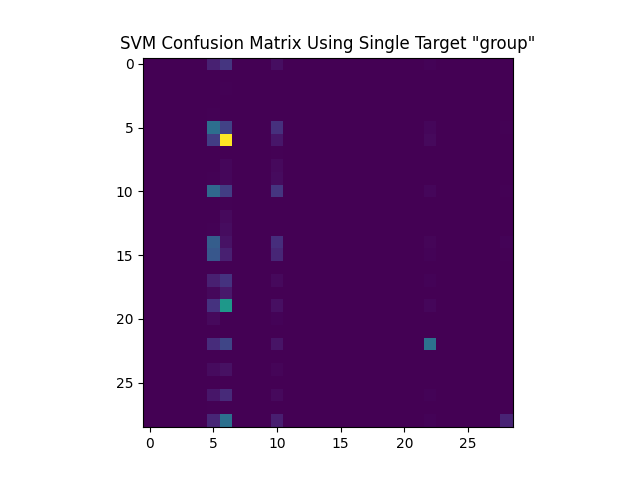}
    \captionof{figure}{SVM – Group}
  \end{minipage}\hfill
  \begin{minipage}[t]{0.32\textwidth}\centering
    \includegraphics[width=\linewidth]{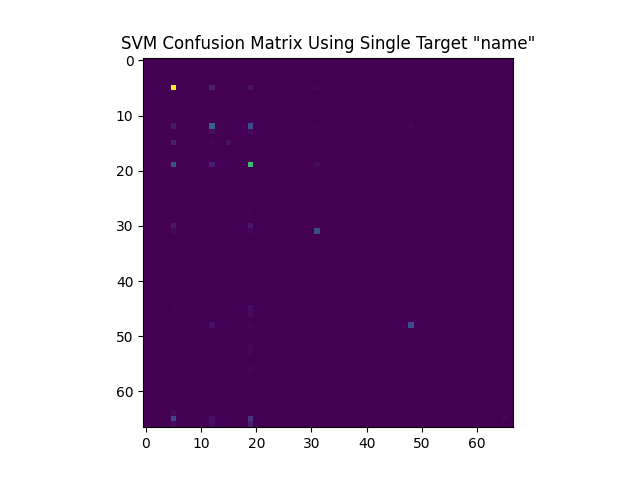}
    \captionof{figure}{SVM – Name}
  \end{minipage}\hfill
  \begin{minipage}[t]{0.32\textwidth}\centering
    \includegraphics[width=\linewidth]{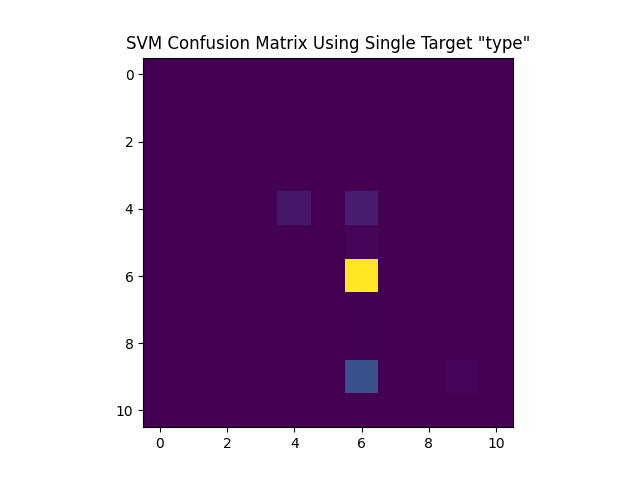}
    \captionof{figure}{SVM – Type}
  \end{minipage}
\end{figure}

\begin{figure}[H]
  \centering
  \begin{minipage}[t]{0.32\textwidth}\centering
    \includegraphics[width=\linewidth]{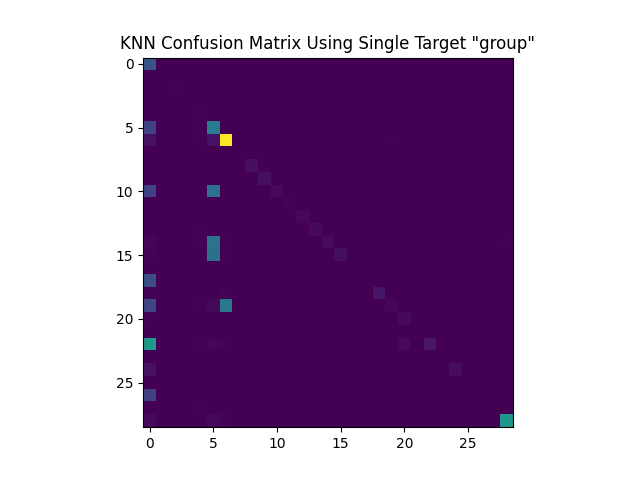}
    \captionof{figure}{KNN – Group}
  \end{minipage}\hfill
  \begin{minipage}[t]{0.32\textwidth}\centering
    \includegraphics[width=\linewidth]{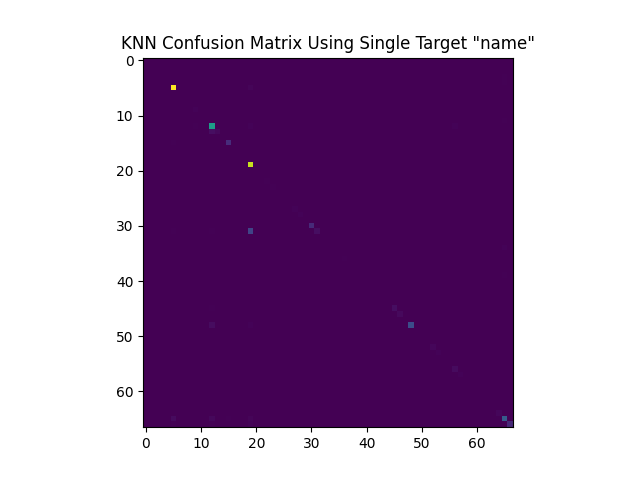}
    \captionof{figure}{KNN – Name}
  \end{minipage}\hfill
  \begin{minipage}[t]{0.32\textwidth}\centering
    \includegraphics[width=\linewidth]{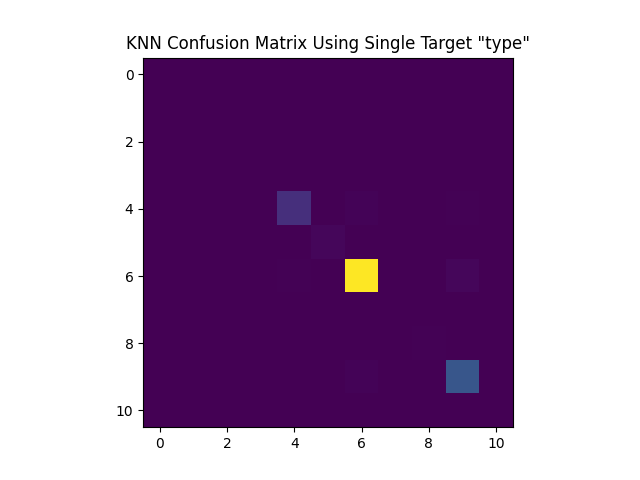}
    \captionof{figure}{KNN – Type}
  \end{minipage}
\end{figure}

\begin{figure}[H]
  \centering
  \begin{minipage}[t]{0.32\textwidth}\centering
    \includegraphics[width=\linewidth]{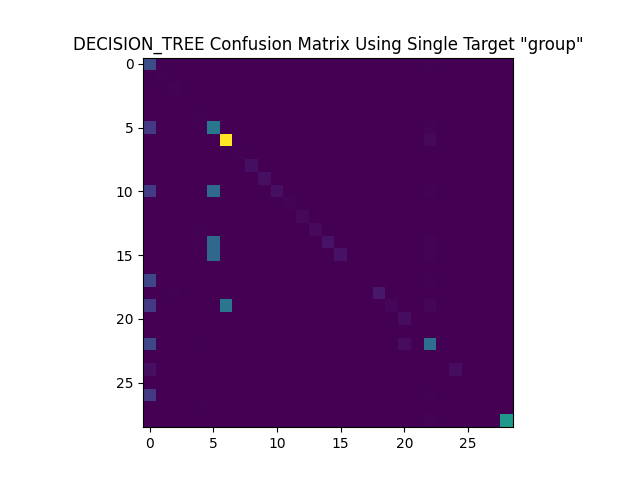}
    \captionof{figure}{Decision Tree – Group}
  \end{minipage}\hfill
  \begin{minipage}[t]{0.32\textwidth}\centering
    \includegraphics[width=\linewidth]{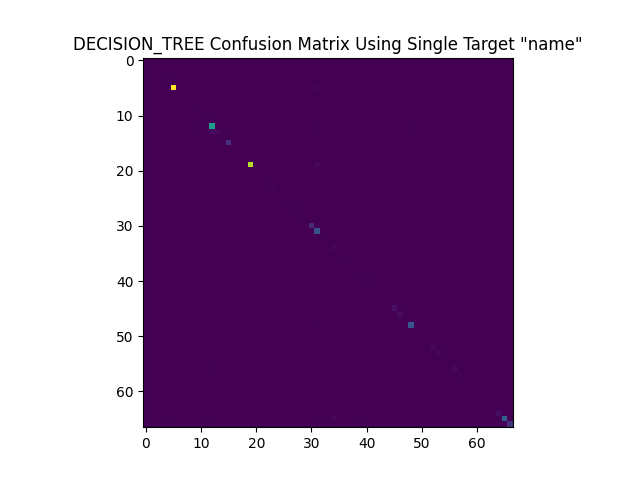}
    \captionof{figure}{Decision Tree – Name}
  \end{minipage}\hfill
  \begin{minipage}[t]{0.32\textwidth}\centering
    \includegraphics[width=\linewidth]{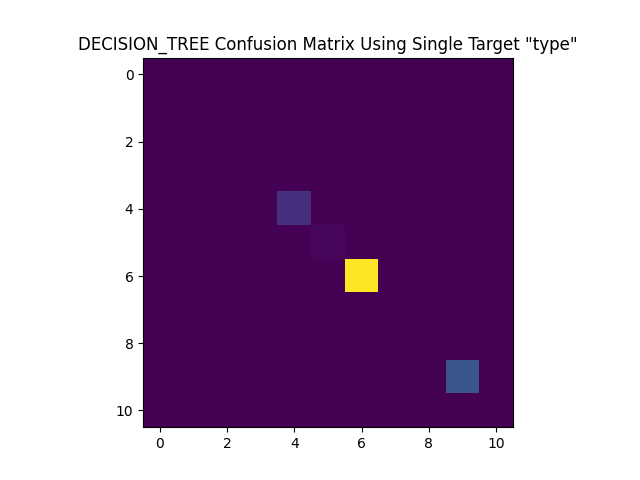}
    \captionof{figure}{Decision Tree – Type}
  \end{minipage}
\end{figure}

\begin{figure}[H]
  \centering
  \begin{minipage}[t]{0.32\textwidth}\centering
    \includegraphics[width=\linewidth]{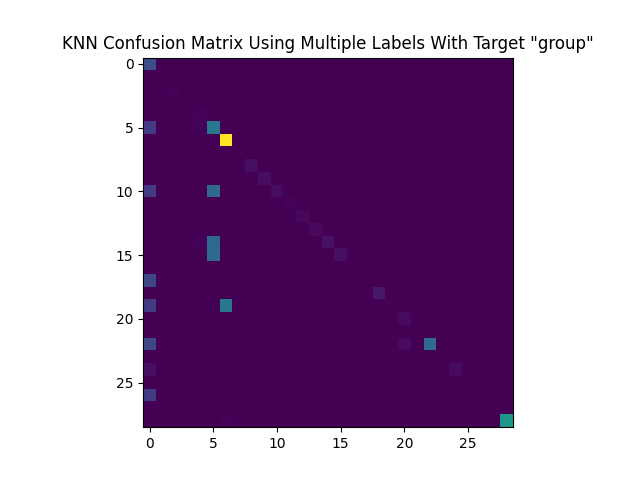}
    \captionof{figure}{KNN – Group}
  \end{minipage}\hfill
  \begin{minipage}[t]{0.32\textwidth}\centering
    \includegraphics[width=\linewidth]{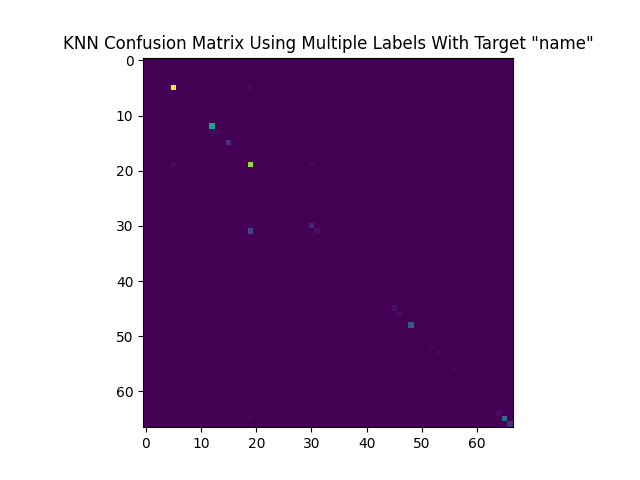}
    \captionof{figure}{KNN – Name}
  \end{minipage}\hfill
  \begin{minipage}[t]{0.32\textwidth}\centering
    \includegraphics[width=\linewidth]{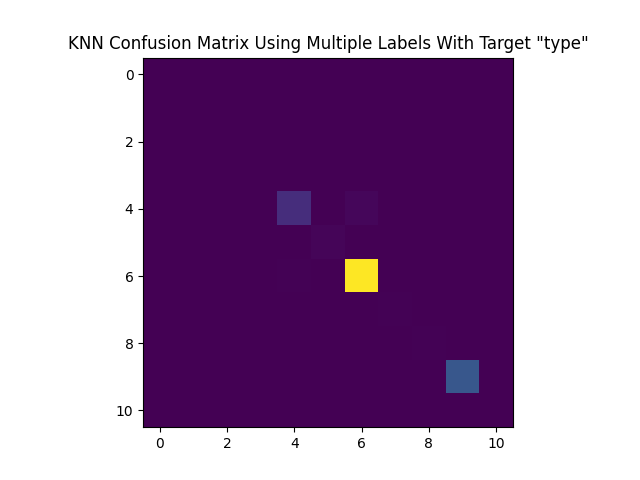}
    \captionof{figure}{KNN – Type}
  \end{minipage}
\end{figure}

\begin{figure}[H]
  \centering
  \begin{minipage}[t]{0.32\textwidth}\centering
    \includegraphics[width=\linewidth]{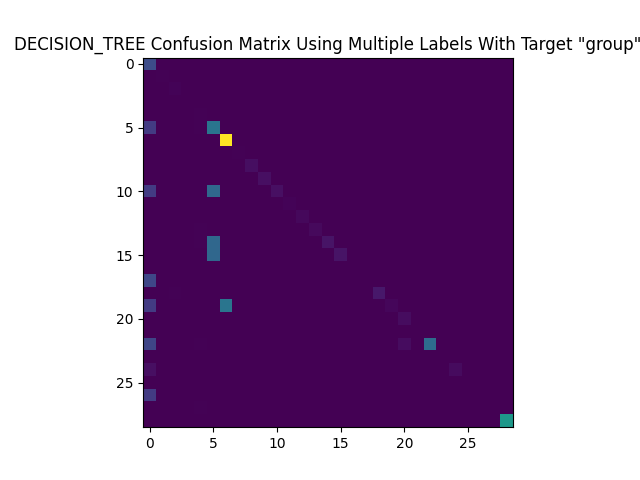}
    \captionof{figure}{Decision Tree – Group}
  \end{minipage}\hfill
  \begin{minipage}[t]{0.32\textwidth}\centering
    \includegraphics[width=\linewidth]{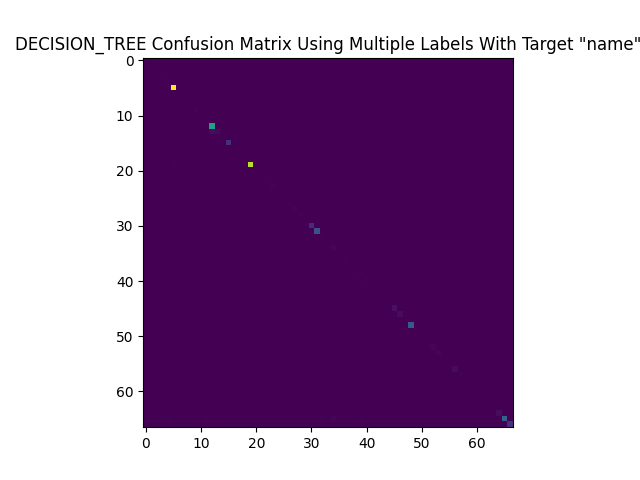}
    \captionof{figure}{Decision Tree – Name}
  \end{minipage}\hfill
  \begin{minipage}[t]{0.32\textwidth}\centering
    \includegraphics[width=\linewidth]{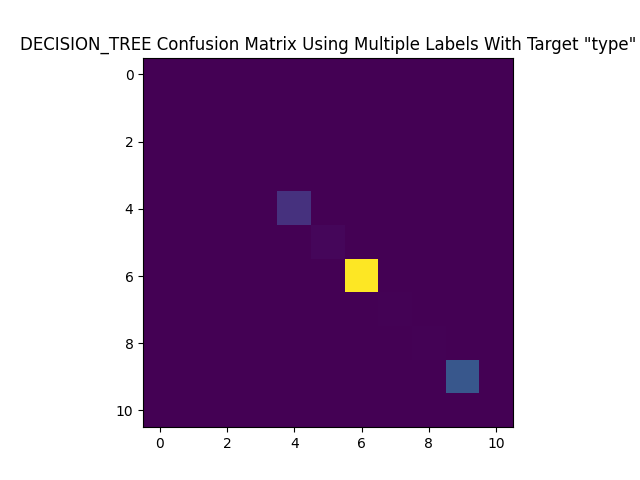}
    \captionof{figure}{Decision Tree – Type}
  \end{minipage}
\end{figure}

\FloatBarrier

\section{Appendix: 1-Gram Performance Visualizations}

\subsection*{Classifier Performance Visualizations}

\begin{figure}[H]
  \centering
  \begin{minipage}[t]{0.48\textwidth}
    \centering
    \includegraphics[width=\linewidth]{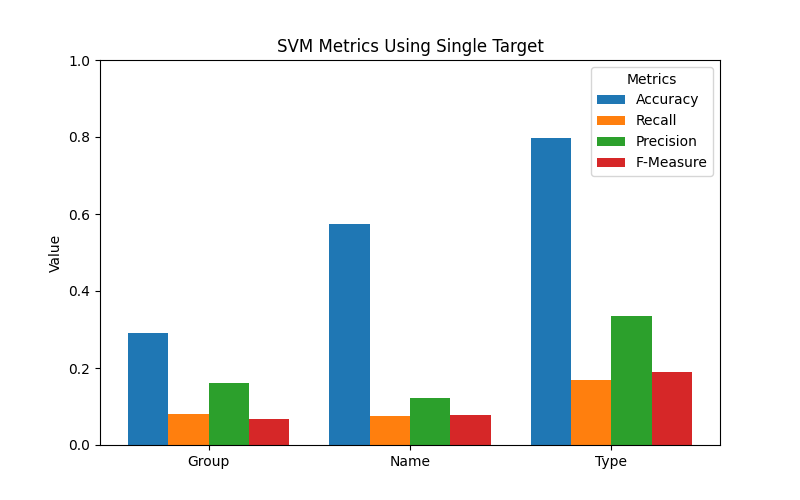}
    \captionof{figure}{SVM – Single Target}
  \end{minipage}\hfill
  \begin{minipage}[t]{0.48\textwidth}
    \centering
    \includegraphics[width=\linewidth]{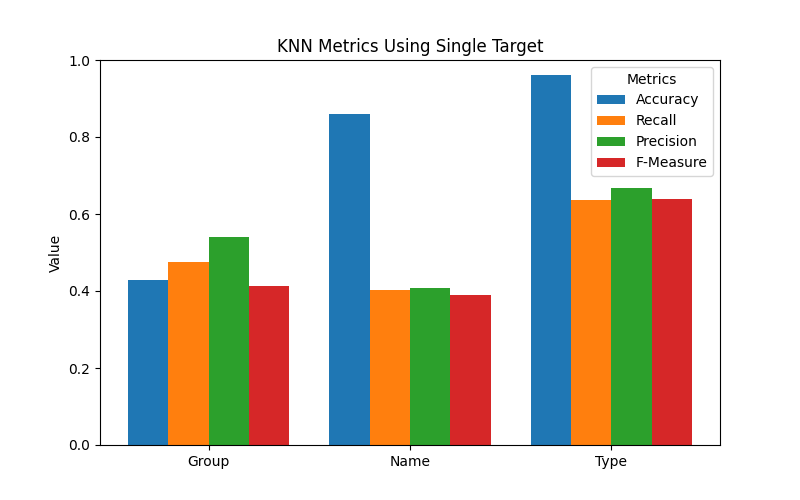}
    \captionof{figure}{KNN – Single Target}
  \end{minipage}
\end{figure}

\begin{figure}[H]
  \centering
  \begin{minipage}[t]{0.48\textwidth}
    \centering
    \includegraphics[width=\linewidth]{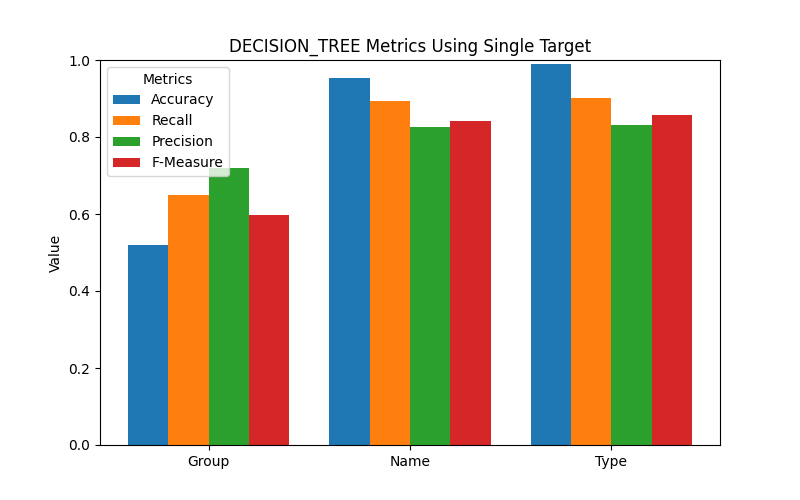}
    \captionof{figure}{Decision Tree – Single Target}
  \end{minipage}\hfill
  \begin{minipage}[t]{0.48\textwidth}
    \centering
    \includegraphics[width=\linewidth]{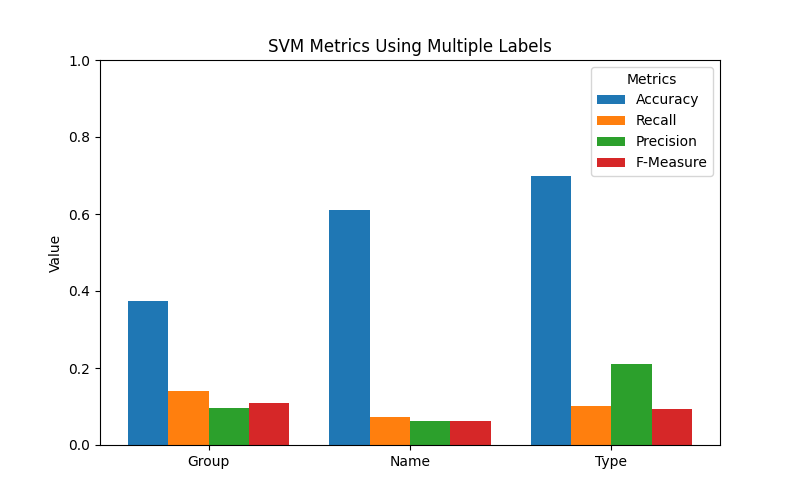}
    \captionof{figure}{SVM – Multi-Label}
  \end{minipage}
\end{figure}

\begin{figure}[H]
  \centering
  \begin{minipage}[t]{0.48\textwidth}
    \centering
    \includegraphics[width=\linewidth]{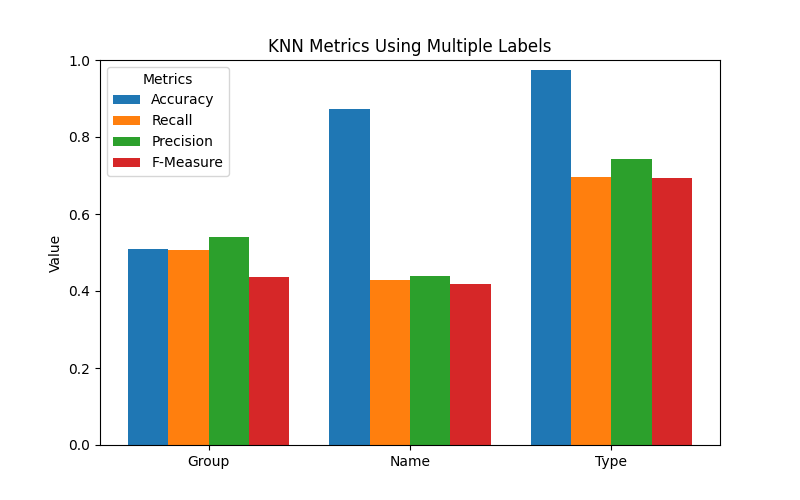}
    \captionof{figure}{KNN – Multi-Label}
  \end{minipage}\hfill
  \begin{minipage}[t]{0.48\textwidth}
    \centering
    \includegraphics[width=\linewidth]{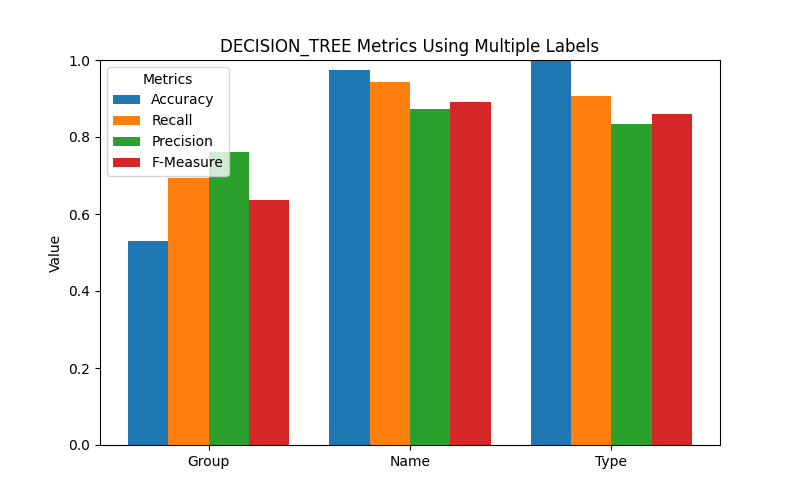}
    \captionof{figure}{Decision Tree – Multi-Label}
  \end{minipage}
\end{figure}

\begin{figure}[H]
  \centering
  \begin{minipage}[t]{0.48\textwidth}
    \centering
    \includegraphics[width=\linewidth]{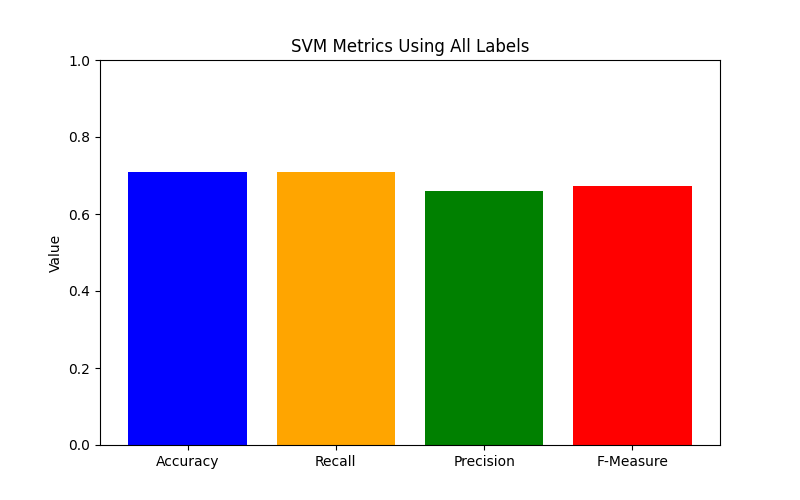}
    \captionof{figure}{SVM – All Labels}
  \end{minipage}\hfill
  \begin{minipage}[t]{0.48\textwidth}
    \centering
    \includegraphics[width=\linewidth]{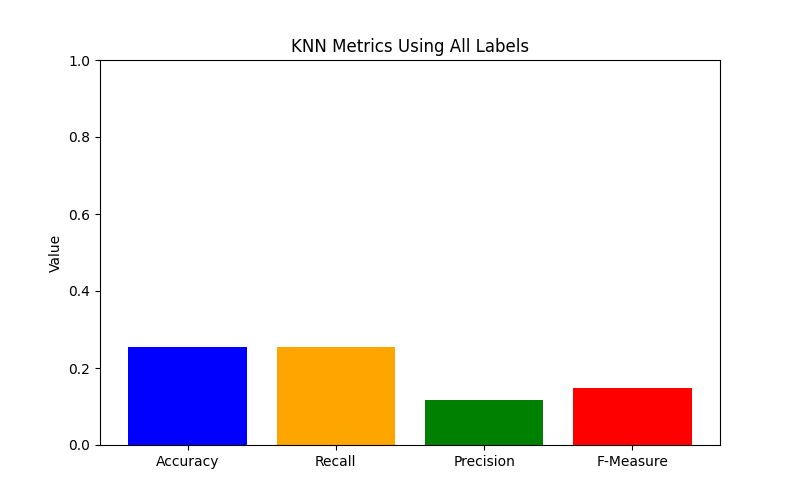}
    \captionof{figure}{KNN – All Labels}
  \end{minipage}
\end{figure}

\begin{figure}[H]
  \centering
  \includegraphics[width=0.48\textwidth]{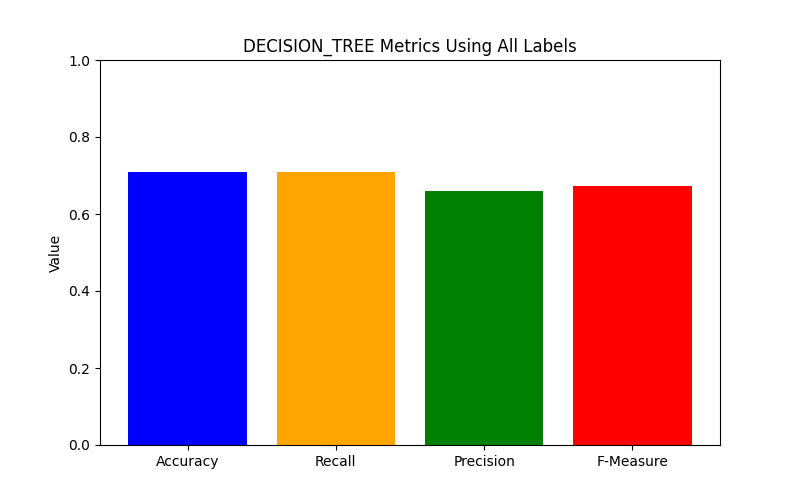}
  \caption{Decision Tree – All Labels}
\end{figure}

\FloatBarrier

\subsection*{Confusion Matrices}

\begin{figure}[H]
\centering
\begin{minipage}[t]{0.32\textwidth}\centering
    \includegraphics[width=\linewidth]{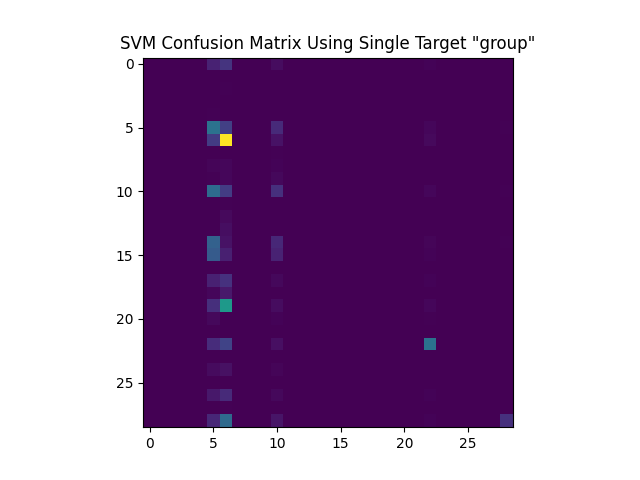}
    \captionof{figure}{SVM – Group}
\end{minipage}\hfill
  \begin{minipage}[t]{0.32\textwidth}\centering
    \includegraphics[width=\linewidth]{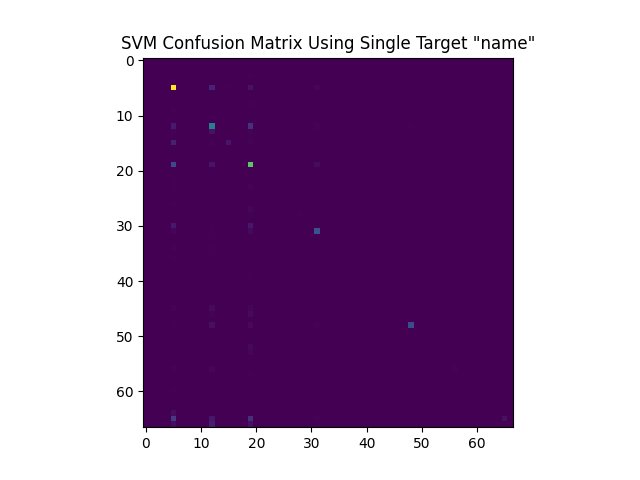}
    \captionof{figure}{SVM – Name}
\end{minipage}\hfill
  \begin{minipage}[t]{0.32\textwidth}\centering
    \includegraphics[width=\linewidth]{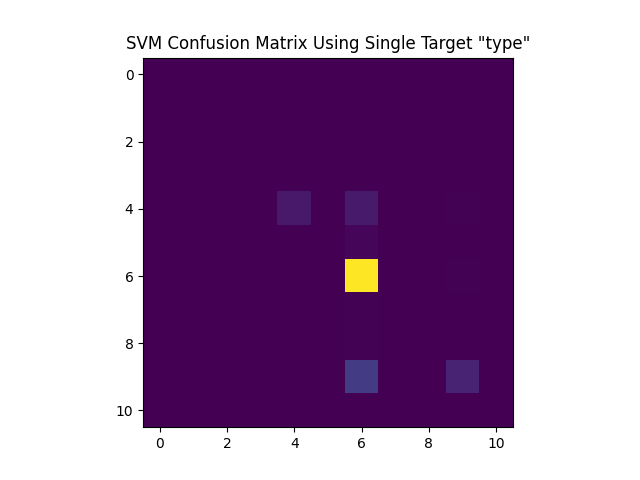}
    \captionof{figure}{SVM – Type}
\end{minipage}
\end{figure}

\begin{figure}[H]
\centering
\begin{minipage}[t]{0.32\textwidth}\centering
    \includegraphics[width=\linewidth]{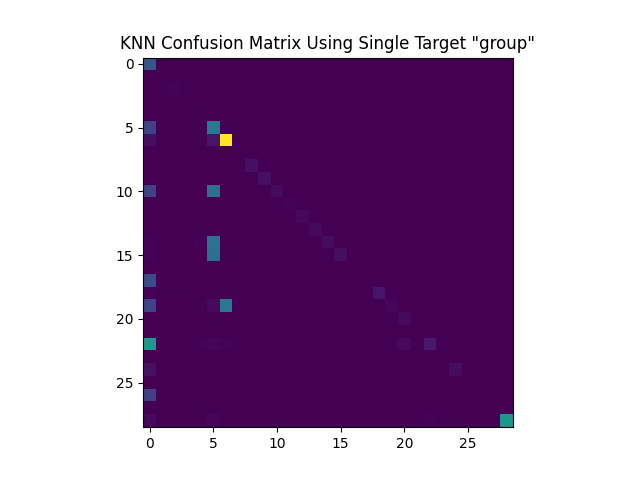}
    \captionof{figure}{KNN – Group}
\end{minipage}\hfill
  \begin{minipage}[t]{0.32\textwidth}\centering
    \includegraphics[width=\linewidth]{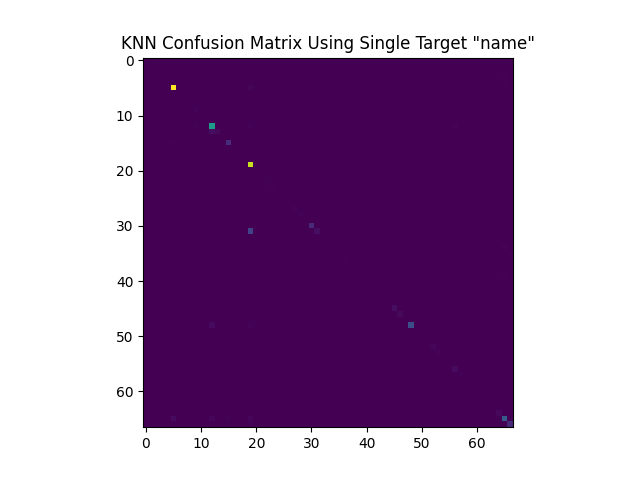}
    \captionof{figure}{KNN – Name}
\end{minipage}\hfill
  \begin{minipage}[t]{0.32\textwidth}\centering
    \includegraphics[width=\linewidth]{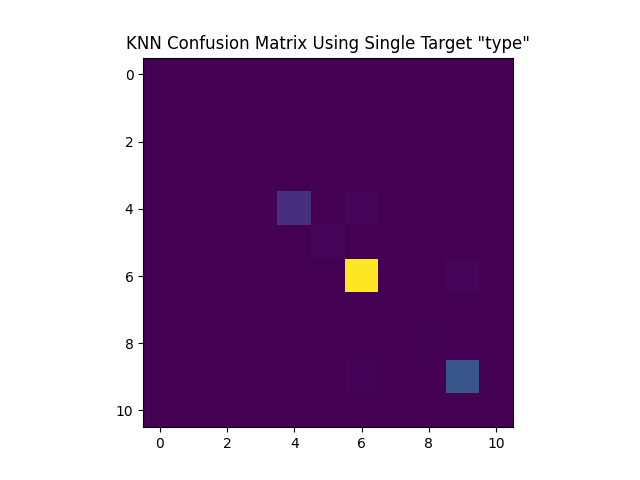}
    \captionof{figure}{KNN – Type}
\end{minipage}
\end{figure}

\begin{figure}[H]
\centering
\begin{minipage}[t]{0.32\textwidth}\centering
    \includegraphics[width=\linewidth]{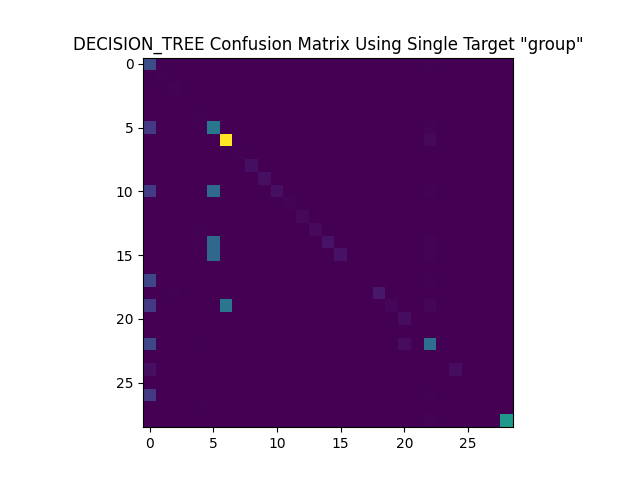}
    \captionof{figure}{Decision Tree – Group}
\end{minipage}\hfill
  \begin{minipage}[t]{0.32\textwidth}\centering
    \includegraphics[width=\linewidth]{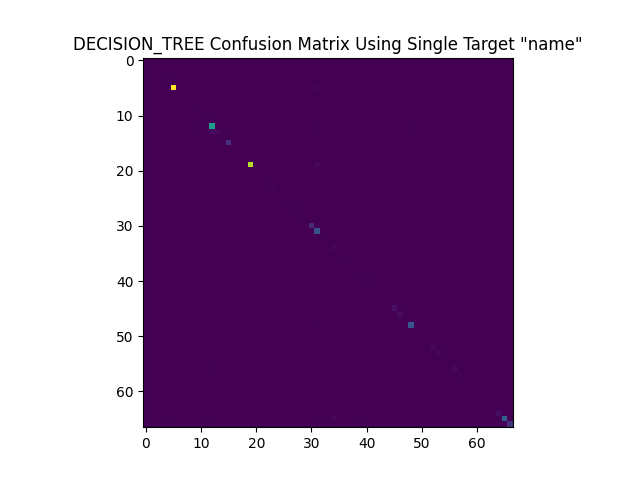}
    \captionof{figure}{Decision Tree – Name}
\end{minipage}\hfill
  \begin{minipage}[t]{0.32\textwidth}\centering
    \includegraphics[width=\linewidth]{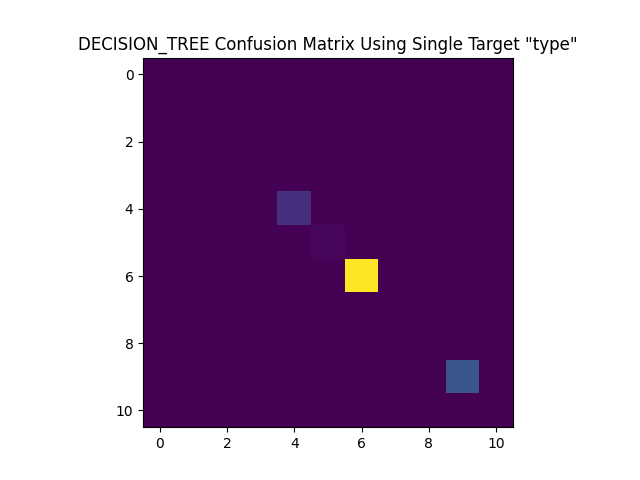}
    \captionof{figure}{Decision Tree – Type}
\end{minipage}
\end{figure}

\begin{figure}[H]
\centering
\begin{minipage}[t]{0.32\textwidth}\centering
    \includegraphics[width=\linewidth]{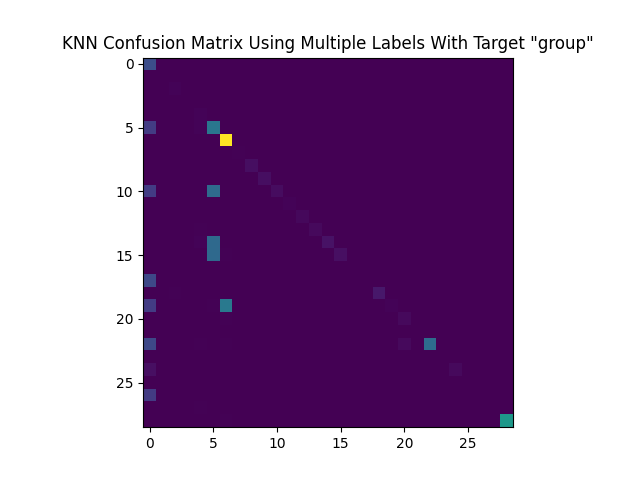}
    \captionof{figure}{KNN – Group}
\end{minipage}\hfill
  \begin{minipage}[t]{0.32\textwidth}\centering
    \includegraphics[width=\linewidth]{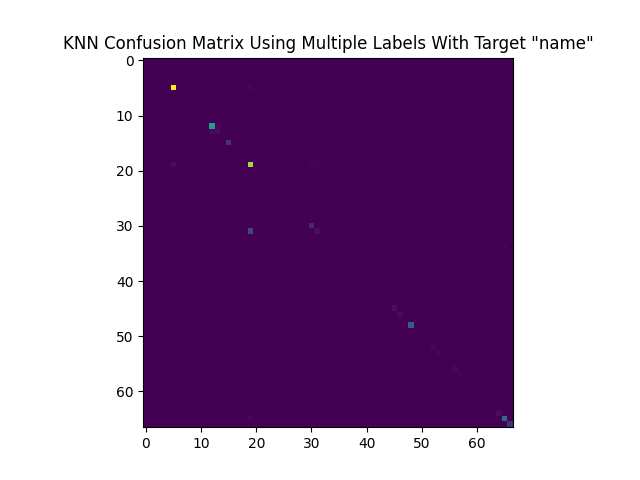}
    \captionof{figure}{KNN – Name}
\end{minipage}\hfill
  \begin{minipage}[t]{0.32\textwidth}\centering
    \includegraphics[width=\linewidth]{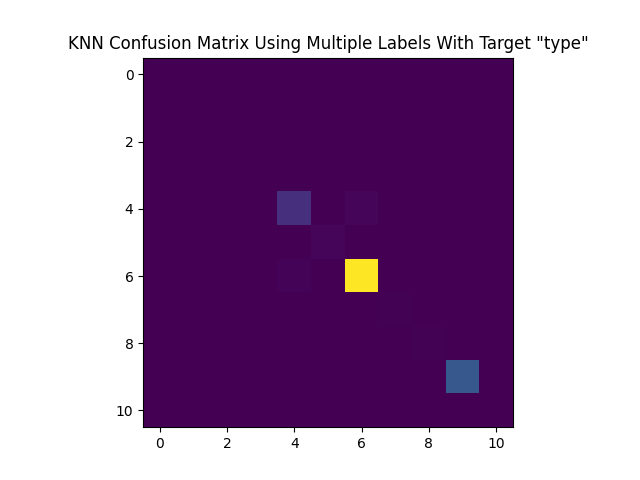}
    \captionof{figure}{KNN – Type}
\end{minipage}
\end{figure}

\begin{figure}[H]
\centering
\begin{minipage}[t]{0.32\textwidth}\centering
    \includegraphics[width=\linewidth]{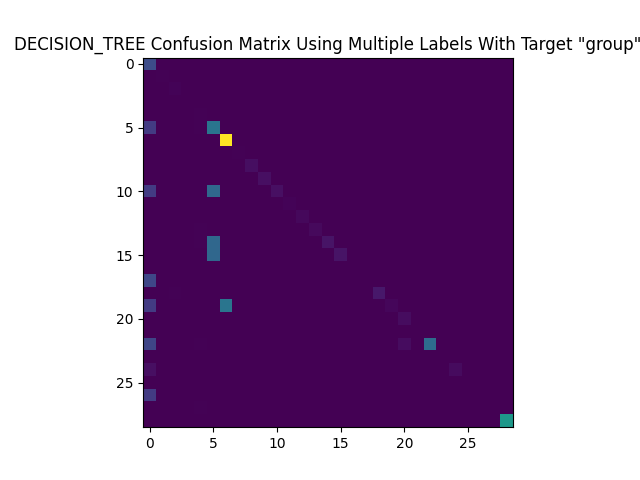}
    \captionof{figure}{Decision Tree – Group}
\end{minipage}\hfill
  \begin{minipage}[t]{0.32\textwidth}\centering
    \includegraphics[width=\linewidth]{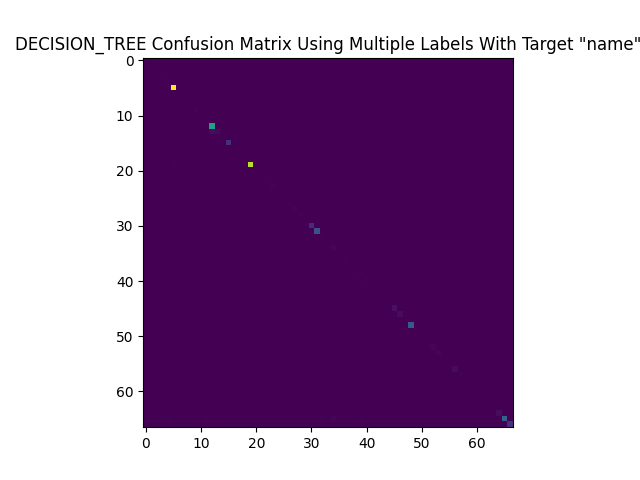}
    \captionof{figure}{Decision Tree – Name}
\end{minipage}\hfill
  \begin{minipage}[t]{0.32\textwidth}\centering
    \includegraphics[width=\linewidth]{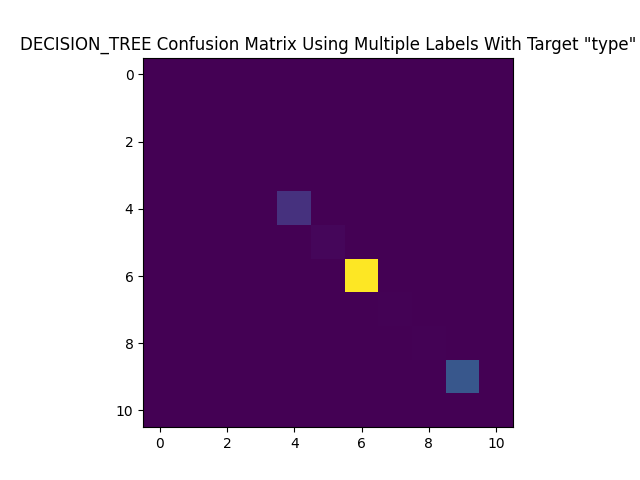}
    \captionof{figure}{Decision Tree – Type}
\end{minipage}
\end{figure}

\FloatBarrier

\end{document}